# A Multi-Entity Input Output (MEIO) Approach to Sustainability — Water-Energy-GHG (WEG) Footprint Statements in Use Cases from Auto and Telco Industries

Reza Farrahi Moghaddam[a,b,*], Fereydoun Farrahi Moghaddam[a,b], Mohamed Cheriet[a,b]

[a]*Synchromedia Lab, École de technologie supérieure (ETS), University of Quebec (UduQ), Montreal, QC, Canada*
[b]*CIRODD: P.O.Box 6079 Downtown, Montreal, QC, Canada, H3C 3A7*

## Abstract

A new Input-Output model, called the Multi-Entity Input-Output (MEIO) model, is introduced to estimate the *responsibility* of entities and actors of an ecosystem on the footprint and actions of each other. It assumed that the ecosystem is comprised of end users, service providers, and utilities. The proposed MEIO modeling approach can be seen as a realization of the Everybody-in-the-Loop (EitL) framework, which promotes a sustainable future using behaviors and actions that are aware of their ubiquitous eco-socio-environment impacts. In this vision, the behavioral changes could be initiated by providing all actors with their footprint statement, which would be estimated using the MEIO models. First, a naive MEIO model is proposed in the form of a graph of actions and responsibility by considering interactions and goods transfers among the entities and actors along four channels. In this model, the unnormalized responsibility and also the final responsibility among the actors are introduced, and then are used to re-allocate immediate footprint of actors to those who are implicitly responsible. The footprint in the current model is limited to three major impacts: Water usage, Energy consumption, and GHG emissions. The naive model is then generalized to Provider-perspective (P-perspective) and End User-perspective (E-perspective) MEIO models in order to make it more suitable to cases where a large number of end users are served by a provider. The E-perspective modeling approach particularly allows estimating the footprint associated to a specific end user. In two use cases from the auto and Telco industries, it has been observed that the proposed MEIO models are practical and dependable in allocating footprint to the provider and also to the end user, while i) avoiding footprint leakage to the end users and ii) handling the large numbers end users. In addition, it will be shown that the MEIO models could implicitly provide some features of the Scope-3 and LCA approaches. This would make the MEIO models an interesting candidate for integrating and merging various concepts that are otherwise incompatible.

*Corresponding author: Reza Farrahi Moghaddam

*Email addresses:* imriss@ieee.org (Reza Farrahi Moghaddam), farrahi@ieee.org (Fereydoun Farrahi Moghaddam), mohamed.cheriet@etsmtl.ca (Mohamed Cheriet)

*URL:* http://ca.linkedin.com/in/rezafm (Reza Farrahi Moghaddam)

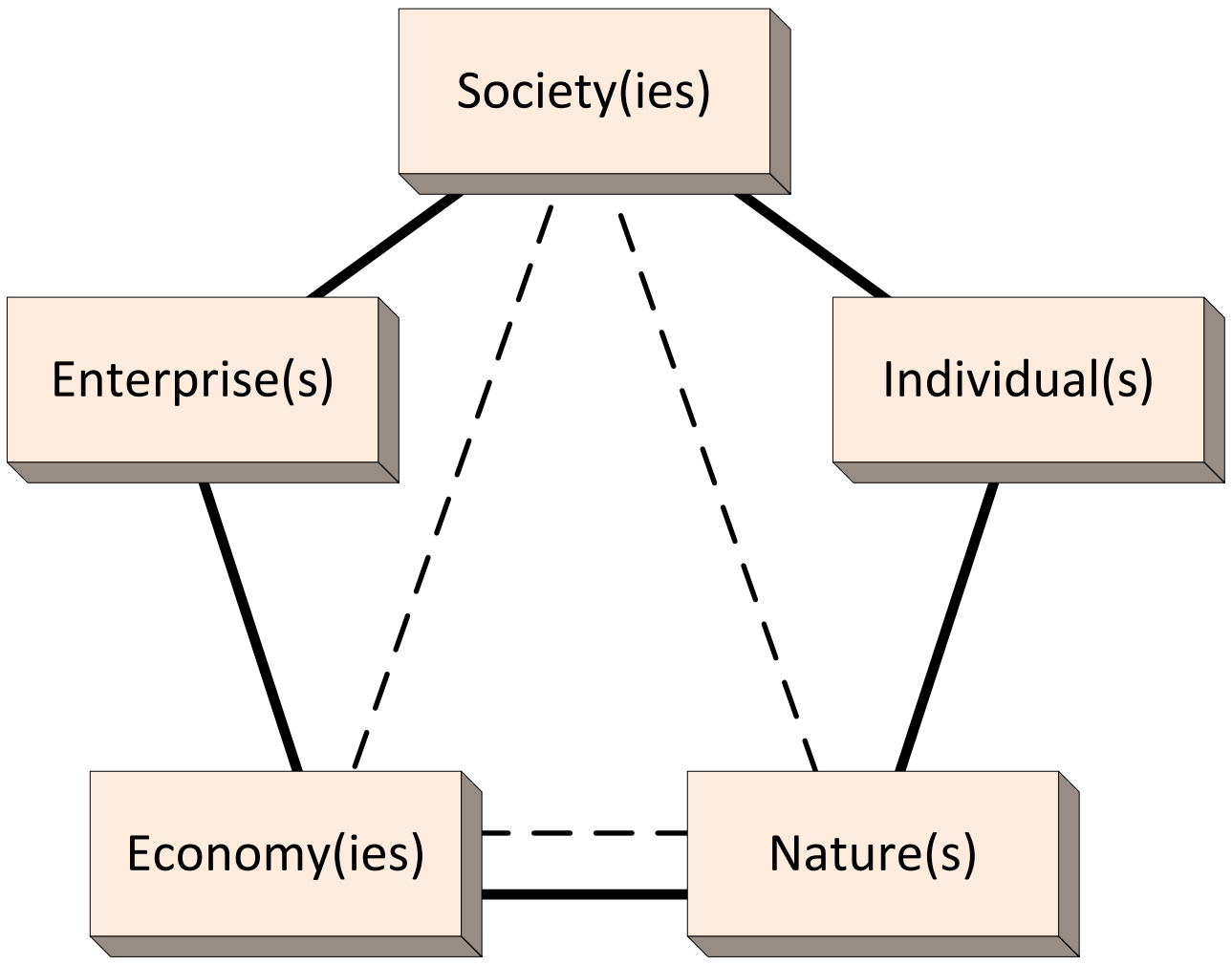

Figure 1: The proposed sustainability model with five elements (from the society perspective: $\mathcal{S}$-perspective). Also, the traditional model is indicated by dashed lines.

## 1. Introduction

Our world is in one of the most critical states along its history. Over-population, over-activity, and over-wasting are some of major trends that could push the world toward an unsustainable state. Awareness in the form of footprint assessment seems to be an effective way toward changing these unsustainable trends. There have been a large amount of research and work to address the complex and challenging problem of estimating and assessing the footprint of various actions and activities (Arushanyan et al., 2014; Farrahi Moghaddam et al., 2014; Seuring and Müller, 2008). These efforts correctly target analysis and assessment of the actors' immediate footprint, and continuously increase the resolution and accuracy of the results at low levels of granularity with respect to actions/actors, spatial and temporal dimensions. Although there are many challenges still ahead of such researches to estimate 'how much' footprint is generated, there are other and more challenging questions to answer: i) 'whom to be charged' with the footprint?, and ii) 'How to charge' them in order to avoid any possible defensive reactions while engaging them toward positive behavior changes? The purpose of this paper is to shed light on these critical but complex aspects of accountability and *responsibility* with respect to others' action, and we will mainly focus on the first question of who is responsible. In our complex and entangled global world, the actual responsible entity for an action could be spatially far from both the actor and the action, which are connected through a complex network of intermediate interactions. Therefore, not only blind charging of the immediate actors could be ineffective if not wrong, a formulation is also required that is capable handle complex interactive ecosystems of big and small actors in order to estimate the final responsibility and footprint of every actor. The proposed MEIO model in this paper is a response to this requirement.

In many service ecosystems, such as those related Auto, ICT, Telco services, there is a high degree of variability and difference in terms of the interests and also capabilities of actors, who can range from individuals, to enterprises, and to societies, as shown in the Sustainability Pentagon (Farrahi

Moghaddam et al., 2014) in Figure 1.[1] This modified sustainability model suggests that adaptive and actor-aware models are required in order to avoid any *responsibility leakage* from one class of actors, such as big enterprises, to another class, such as small end users. The models should also be aware of the large number of actors in a specific class, which could otherwise practically diminish the footprint of every individual actor. For example, in industries such as Telco, simple direct dividend share of the calculated footprint of operation, even when obtained using Life Cycle Assessment (LCA) methodologies or Scope-3 assessments (Downie and Stubbs, 2012, 2013; Huang et al., 2009), among the subscribers could be so negligible that it would not ignite neither any adjustment in the behavior of a subscriber nor any modification in the Business as Usual (BAU) of the Telco provider. At the same time, both the subscribers and the Telcos hold a considerable *responsibility* because otherwise in the absence of their actions there would no footprint at all.

This supports our approach to look at such complex set of actors and actions in the form of a heterogeneous *ecosystem*[2] of *selfish* but *smart* actors who act, react, or inact according to their understanding of the *environment* that they *observe* within the ecosystem (Farrahi Moghaddam et al., 2012b). This is our motivation to propose the Everybody-in-the-Loop (EitL) framework of sustainability, as a generalization of the User-in-the-Loop (UitL) concept (Schoenen et al., 2012, 2011). The EitL targets both *expanding* the observable environment of every actor and also charging everybody with its actual footprint to promote self-motivated, positive changes in actions and behaviors of these *smart* actors. The EitL framework at the same time requires to protect the end users from being charged with footprint beyond their actual *responsibility.* More discussion on the EitL framework is provided in the Appendix A.

In order to realize the EitL framework in the form of a model with tractable results, we here introduce the action-aware Multi-Entity Input Output (MEIO) model, which has some similarities to the Multi-Region Input Output (MRIO) models. In the MEIO models, the entities are the actors of the ecosystem who interact with each other without any requirement that they are associated to a specific region. In addition, the MEIO modeling relaxes the requirement of existence of an input-output relation among the actors by directly considering the actions and interactions that took placed in a certain time interval. More details are provided in section 3.

Before proceeding, we would like to highlight a challenge ahead of all assessments and analyses that we call the *curse of big numbers.* While aggregation and consolidation are common practices to reach an *unique* conclusion that can be then used to initiate proper actions, the dominance of big numbers would shadow the *significance* of the small ones. A similar phenomenon is the *winner's curse* in some of human activities (Athias and Nuñez, 2006, 2008; Nuñez, 2007). Another challenge would be *limited* range of footprint's impact, for example, in the case of some types of air pollution, considering nowadays global interactions among actors that are spread across multiple fundamentally-different regions (Farrahi Moghaddam et al., 2014). The MEIO model could provide an alternative answer to these challenge because it assesses *every* actor's *responsibility* regardless of their size and place. This low-level of granularity at the level of actors opens a lot of opportunities for proper decision makings and sustainability measures.

[1]The proposed modified sustainability model consists of five components compared to the three-component classical model (Glavič and Lukman, 2007).

[2]As a generalization of the original *natural* ecosystem definition (Biggs et al., 2012).

As an example to illustrate the importance of the EitL framework, and its implementation in the form of the MEIO model, we will consider a use case from the Auto industry. Although the details are provided in section 4.1, it is worth mentioning here that the volunteer Scope-3 analysis performed by the automaker of this use case (Honda Motor Co. Ltd., 2012) to include the footprint associated to their buyers' use of cars has resulted in a extra *self-overcharging* of 106.77% compared to what we will calculate for them using the MEIO model. In particular, in section 4.1, the MEIO model will show that Honda automaker is responsible for a share of 113.37 $MtCO_2e$ (48.37%) footprint of their Fiscal Year (FY) 2011 buyers' footprint, while a Scope-3 analysis would assign them the whole footprint of 234.4 $MtCO_2e$ generated by these buyers in 15 years. Although taking *all* the responsibility would be seemed to be harmless if not positive, it could have its own drawbacks considering the aforementioned curse of big numbers. An immediate side effect would be shadowing or downplaying many possible *internal* actions, such as avoiding peak hours operation to help the electricity grid do not fire up its non-green reserved power plants that the enterprise would consider to reduce its footprint, simply because they would become smaller than the "threshold" (Busch, 2011).

The proposed ecosystem picture is along the fact that the most of providers' action is in response to service requests from the end users. This means that, considering the complexity and competitiveness involved, the service providers could implicitly promote more number of requests by introducing new services or interactions. Therefore, assigning all the responsibility to the end users would be unfair and also ineffective. In addition, it has been observed that rigid regulations not only cannot solve a problem, they may also skew the fairness of the market and provide ways of abuse for certain providers, especially for those providers that are not within the regulation's jurisdiction. Although mechanisms, such as border adjustment, could be used for both implementing the regulations and also preserving the competitiveness (Farrahi Moghaddam et al., 2013, 2014; Kuik and Hofkes, 2010), they could not be 'zoomed' down to the granularity level required in the case of interactions among enterprises and individual, such as those involved in Auto or Telco industries. As mentioned before, it this work, we introduce the transparent everybody-in-the-loop (EitL) framework to estimate the footprint of every actor even with presence of an inhomogeneous degree of involvement from each actor. The primary outcome of this framework would be the footprint statements of the end users. These statements could be then used to issue charges with a two-fold impact: i) Encouraging the end users to adjust their service-request patterns and also to drop unnecessary requests, and ii) Guiding the end users to choose or switch to those providers that generate less footprint. The latter is actually an implicit mechanism to encourage the providers to adapt more sustainable operation in order to prevent mass losses of users. These mechanisms are recommended to be executed by bodies other than the providers themselves, and also it is a natural choice to redistribute the generated revenue among all the end users, or among all the individual residing in the associated society. The secondary outcome of the EitL framework would be the footprint statements of the providers. Although these statements may not be used to charge the providers, they provide a clear picture of their commitment to sustainability, and therefore can influence the loyalty of the end users (Bocken et al., 2014; Boons and Lüdeke-Freund, 2013).

Presence of individual end users in the ecosystemic picture of the EitL framework highlights another aspect, i.e., the need for 'behavioral' modeling. In contrast to the big actors that *move* slowly, the individuals could change their mind instantaneously and even in large masses. This indicates a great opportunity to drive the end users and therefore the whole user-provider-society ecosystem toward a more sustainable state. At the same time, this requires deep considerations in order to reduce the associated risks. We will discuss this in Appendix A. In the EitL framework, we suggest to leave it to every individual to cognitively plan and move toward sustainability of the world and also their own "branch."

The paper is organized as follows. In section 2, some of the mathematical notation used to describe the graph of the interactions among actors are introduced. Then, in section 3 using a graph theory approach, the proposed MEIO modeling framework is presented in three version: naive, provider perspective, and end user perspective. In section 3.1.1, an illustrative example of an ecosystem of four imaginary actors is provided to show how the naive MEIO model works. Furthermore, in the Experimental Results section, section 4, two real use cases from Auto and ICT industries are presented to show the capabilities of the MEIO models and also to highlight the fundamental differences among them and other approaches. Finally, the conclusions and some future prospects are discussed in section 5. In addition, in two appendices, the relations among actors' behavior and sustainability of an ecosystem is discussed considering the proposed modified sustainability model, and also the concept of gauge econons, which is used to define the interactions in the MEIO models, is presented.

It is worth noting that the use cases provided in section 4 are subject to improvement and should not be considered by any means as complete analyses of the associated enterprises. These use cases are merely provided as examples to show *how* the MEIO models could be used. The selection of these enterprises for our use cases could be implicitly considered as our acknowledgment of their effort toward sustainability without downplaying the others.

## 2. Notations and Definitions

In this section, some of the graph theory notations used from here on are presented. The rest of notations is provided in section 3 where the MEIO models are introduced.

- $G(V, E, w)$: A directed graph with vertex[3] set $V$ of $n = ||V||$ nodes along with the set $E$ of its $m$ active edges and their associated weights $w$. A vertex is denoted either by $v_i$ or by its index $i$: $v_i \in V$. We may also causally use $i \in V$ as a short notation. A typical directed edge is shown by $e_{i,j}$, where $j$ and $i$ are the the start and end vertexes. Also, $w_{j\to i}(= w\,(e_{i,j}) > 1)$ denotes the associated positive weight of the edge $e_{i,j}$. We assume $w > 0$, in order to keep the weights of the associated logarithmic graph positive. Please see the definition of the $G^{L}$ and $G^{\mathrm{LR}}$ graphs below.

- $G^{\mathrm{L}}(V, E, w)$: The logarithmic graph associated with a graph $G(V, E, w)$: $G^{\mathrm{L}}(V, E, w) = G(V, E, w^{\mathrm{L}}) = G(V, E, \log w)$. In other words, the weights of $G^{\mathrm{L}}$ are the logarithm of the weights of the original graph $G$: $w^{\mathrm{L}}_{j\to i} = \log w_{j\to i}$.

[3] will be also referred to as object or actor.

- $G^{\mathrm{LR}}(V,E,w)$: The logarithmic-reversed graph associated with a graph $G(V,E,w)$: $G^{\mathrm{LR}}(V,E,w) = G(V,E,w^{\mathrm{LR}})$, where: $w^{\mathrm{LR}}_{j\to i} = \log\left(2\frac{w_{\max}}{w_{j\to i}}\right)$, and $w_{\max}$ is the maximum value of $w$.

- $r^{\beta}_{i,j}$: A *responsibility path* from the node $j$ to the node $i$, which is a connected, directed path sourced from $j$ and sunk at $i$. The vector $\beta = \{v_{\beta,k} | k = 1, \cdots, k_{\max}\}$, which consists of a few node indices, also acts as a counter on all possible responsibility paths from $j$ to $i$, and the set of these paths is denoted $r_{i,j} = \left\{r^{\beta}_{i,j}\right\}_{\beta}$. For the purpose of simplicity in notation, the dependency of $k_{\max}$ on $\beta$ is implicitly understood. The path $r^{\beta}_{i,j}$ is formally defined as follows:

$$r^{\beta}_{i,j} = \left\{v_{\beta,k} | v_{\beta,k} \in V; v_{\beta,1} = j; v_{\beta,k_{\max}} = i; w_{v_{\beta,k}\to v_{\beta,k+1}} \neq 0, \forall k = 1, \cdots, k_{\max}-1\right\}.$$

  To account for single-edge responsibility paths, the $w_{j\to i}$ is also called the *immediate responsibility* of node (actor) $j$ on the node (actor) $i$ if they are immediate neighbors.

- $s_{i,j}$: The *shortest directed path(s)* from the node $j$ to the node $i$. It is defined as: $s_{i,j} = \underset{\beta}{\operatorname{argmin}} \sum_{k=1} \left(w_{v_{\beta,k}\to v_{\beta,k+1}}\right)$. The length of this path is denoted $||s_{i,j}||$.

- $\bar{R}^{\beta}_{i,j}$: The *unnormalized responsibility* of the node $j$ on the node $i$ *along* the responsibility path $r^{\beta}_{i,j}$:

$$\bar{R}^{\beta}_{i,j} = \prod_{k=1}^{k_{\max}-1} w_{v_{\beta,k}\to v_{\beta,k+1}}.$$

- $S_{i,j}$: The *strongest directed path(s)* from the node $j$ to the node $i$. It is defined as: $S_{i,j} = \underset{\beta}{\operatorname{argmax}}\, \bar{R}^{\beta}_{i,j}$.

- $\bar{R}_{i,j}$: The *unnormalized responsibility* of the node (actor) $j$ on the node (actor) $i$, which is defined: $\bar{R}_{i,j} = \max_{\beta}\left\{\bar{R}^{\beta}_{i,j}\right\}_{\beta}$. It can be also imagined as a matrix of unnormalized responsibilities among actors in the form of $\bar{R}$.

- $R_{i,j}$: The *responsibility* of the node (actor) $j$ on the node (actor) $i$, defined by normalizing $\bar{R}$ on the source side:

$$R_{i,j} = \bar{R}_{i,j} \Big/ \left(\sum_{j}^{\prime} \bar{R}_{i,j'}\right).$$

  The reason for performing normalization on the *source* side is that we will use these values to share the footprint of the actor $i$ among the sources.

- $\bar{\lambda}$: An *edge perturbation* vector $\bar{\lambda} = \left(\bar{\lambda}_{\eta}\right)_{\eta=1}^{m}$ of $\lambda$-*perturbation* degree, i.e., $\lambda \leq \bar{\lambda}_{\eta} \leq 1$, $\eta = 1, \cdots, m$ (Bilò et al., 2010). The definition is modified according to the *strongest path* problem of this work in contrast to the *shortest path* problem in (Bilò et al., 2010). The associated perturbed graph $G_{\bar{\lambda}} = G(V,E,w_{\bar{\lambda}})$ is defined by *multiplying* the weights of the original $G(V,E,w)$ by their associated perturbation value from $\bar{\lambda}$.

- $\delta(H)$: The *stability number* of a subgraph $H$ of $G$ with respect to an optimization problem $\mathcal{O}$, such as the strongest path optimization problem. In the short notation, $\delta(H)$ is also denoted by $\sigma_H$. Also, for the purpose of simplicity, the notation $\mathcal{O}$ is omitted from the $\delta(H)$ notation. The $\delta(H)$ is defined as the minimum value that for every edge perturbation $\bar{\lambda}$ of degree $\delta(H)$, $H$ contains an optimal solution for the graph $G_{\bar{\lambda}}$ with respect to $\mathcal{O}$ (Bilò et al., 2010).

- $S^{\delta}_{i,j}$: The strongest path(s) valid under the perturbation of degree $\sigma_{H_{i,j}}$, where $H_{i,j}$ is the subgraph of $G$ induced by all the strongest path(s) from the node $j$ to the node $i$ in the original graph $G$. Formally, we define: $S^{\delta}_{i,j} = \left\{ r^{\beta}_{i,j} \middle| \left\| r^{\beta}_{i,j} \right\| \geq \sigma_{H_{i,j}} \left\| S_{i,j} \right\| \right\}$. In cases where $H_{i,j}$ is only one path, it is considered that $\sigma_{H_{i,j}} = +\infty$.

## 3. The Proposed Multi-entity Input Output (MEIO) framework

Because of the complexity associated to the ecosystems that involve end users, we introduce three versions of the proposed MEIO model. The first one, a naive model, intentionally ignores presence of the end users. In this model, all actors are treated in the same way. Then, two MEIO models, which are aware of presence of the end users, are introduced to estimate the footprint of a provider and also the footprint of an end user while preventing footprint "leakage" to or from the end users. To do so, the end users should be explicitly separated from the rest of the actors. This will be carried out with introduction of the sector *slicing hyper-line* which limits intervisibility of the end users and other actors only to the actions associated to the sector they are interacting on. Also, some discussions on the similarities and differences between MRIO and MEIO models will be presented in section 3.1.3.

### *3.1. The Naive MEIO Model*

Let us consider a world $W$ of entities,[4] which is partitioned into the set of Entities of Interest (EoI) and the Rest of the World (RoW). Furthermore, we assume the RoW is an atomic entity, and this brings us to a modified world $V$ comprised of the RoW entity and the entities in the EoI set. Also, let us assume a specific interval of time $\Delta t$ that we are interested to analyze actions and their associated footprint.

The specific feature of the MEIO models is the way they handle interactions among actors. In contrast to the traditional link-based ('tie'-based) approaches, in which an action represents somehow a link or a tie between participating entities, we consider actions themselves as "objects" in the MEIO models. This picture has been successfully used in particle physics where interaction particles, such as Gauge bosons, play the role of *mediators* of interactions between other particles (Langacker, 2009). We generalize this concept to the 'gauge econons' objects in the MEIO models that represent goods transfer and other interactions among the entities.

We have to consider all asset exchanges (not just money-based ones) in order to prevent any loophole in which footprints are moved and isolated within *designated* entities with minimal money-based exchanges with truly responsible entities. To do so, we consider high-level sectors of the economy. There are three major sectors in economy: Primary (raw material and food), secondary (finished goods), and tertiary (services). However, we separate the banking from the third sector because money and credit exchange represents the main interaction with the end users. Therefore, we have 4 sectors/exchange types to be considered, and they are visualized as four *channels* of interaction and transfer among the entities. Each transfer or exchange channel is considered to be normalized to do not influence the others. Examples of goods from each of these channels, in both physical and virtual forms, are presented in Table 1. For more discussion, please see Appendix B.

[4] Entities could act as actors or inactors, among other possible roles in interactions.

| | Channel | Physical sub-channel | Non-physical sub-channel |
|---|---|---|---|
| 1 | Raw | A ton of Iron | Data |
| 2 | Goods | A TV set | A digital movie |
| 3 | Services | Waste collection | Movie on-demand |
| 4 | Money | Cash | Digital currency |

Table 1: Examples from the four channels considered in this work in both physical and and non-physical sub-channels.

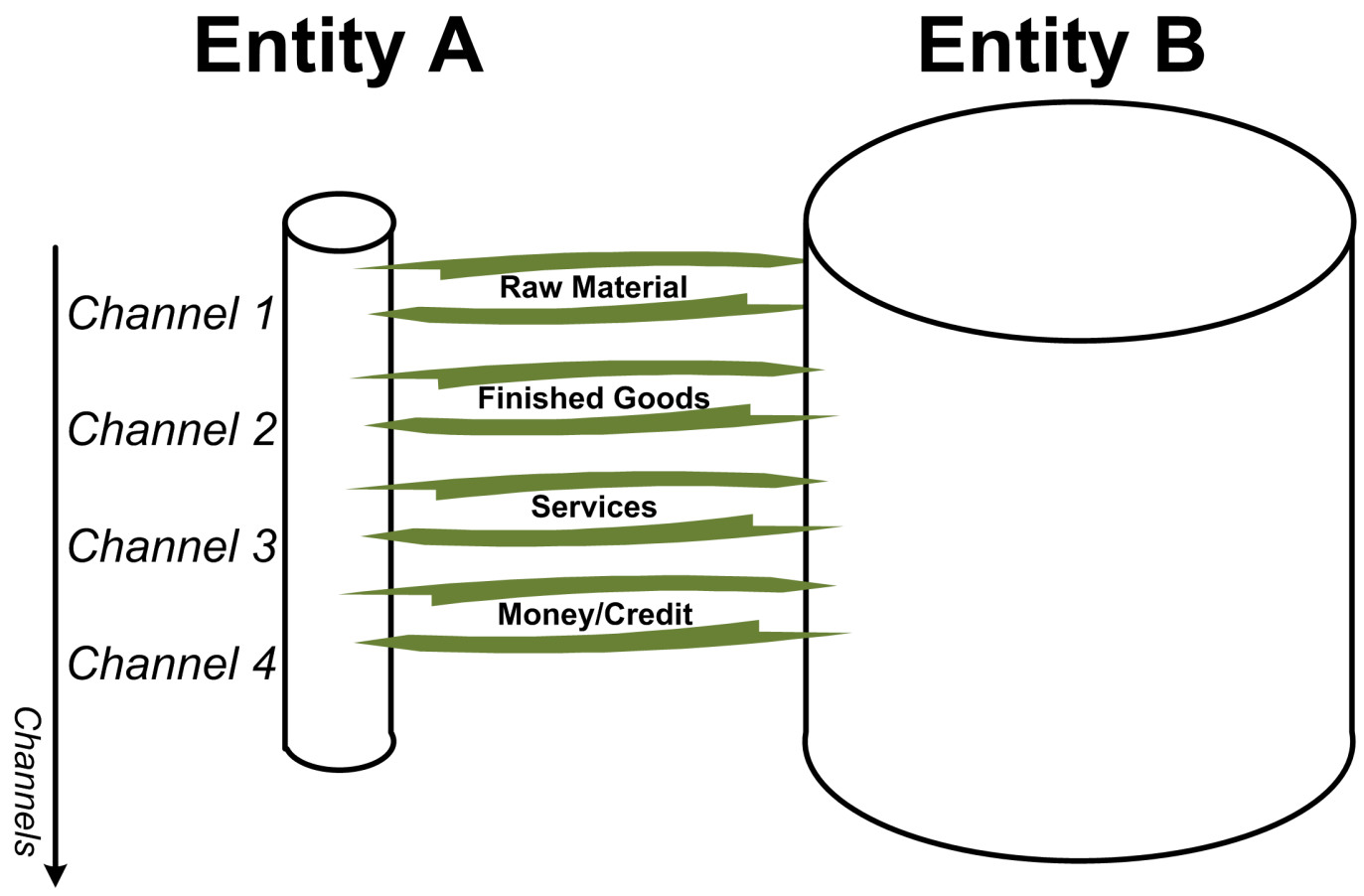

Figure 2: The four channels that are considered to model interactions between entities in the proposed MEIO framework.

To do this, each asset type is considered with an equal weight of one even if it represents a fraction of money exchanges of an entity. We argue that if that asset type is really insignificant in the business model of the entity, they could drop it or simply source it from an entity with lower footprint intensity without affecting their profit. The details of enforcing this requirement is provided below when the edges' weight of the responsibility's graph are calculated.

The set of actions, i.e., the gauge econons, among the entities of EoI within the $\Delta t$, and also those aggregated actions between these entities and the RoW is denoted $\Xi = \{\zeta_\theta\}_\theta$, where $\zeta_\theta$ represents a typical econon. Every action is carried along a channel from the set of possible channels $\mathcal{C}$:

$$\mathcal{C} = \{c_{\text{Raw}}, c_{\text{Goods}}, c_{\text{Services}}, c_{\text{Money}}\} = \{c_{\text{R}}, c_{\text{G}}, c_{\text{S}}, c_{\text{M}}\}, \tag{1}$$

in the current form of the proposed models (please see section Appendix B). Also, the set of *downward* footprint actions of the entities in the EoI on the nature(s) is denoted $\Phi = \left\{\bar{F}_{i,f}\right\}_{i,f}$, where $\bar{F}_{i,f}$ is an immediate footprint action from the entity $i$ on the natural resources in the impact *category* $f$, and

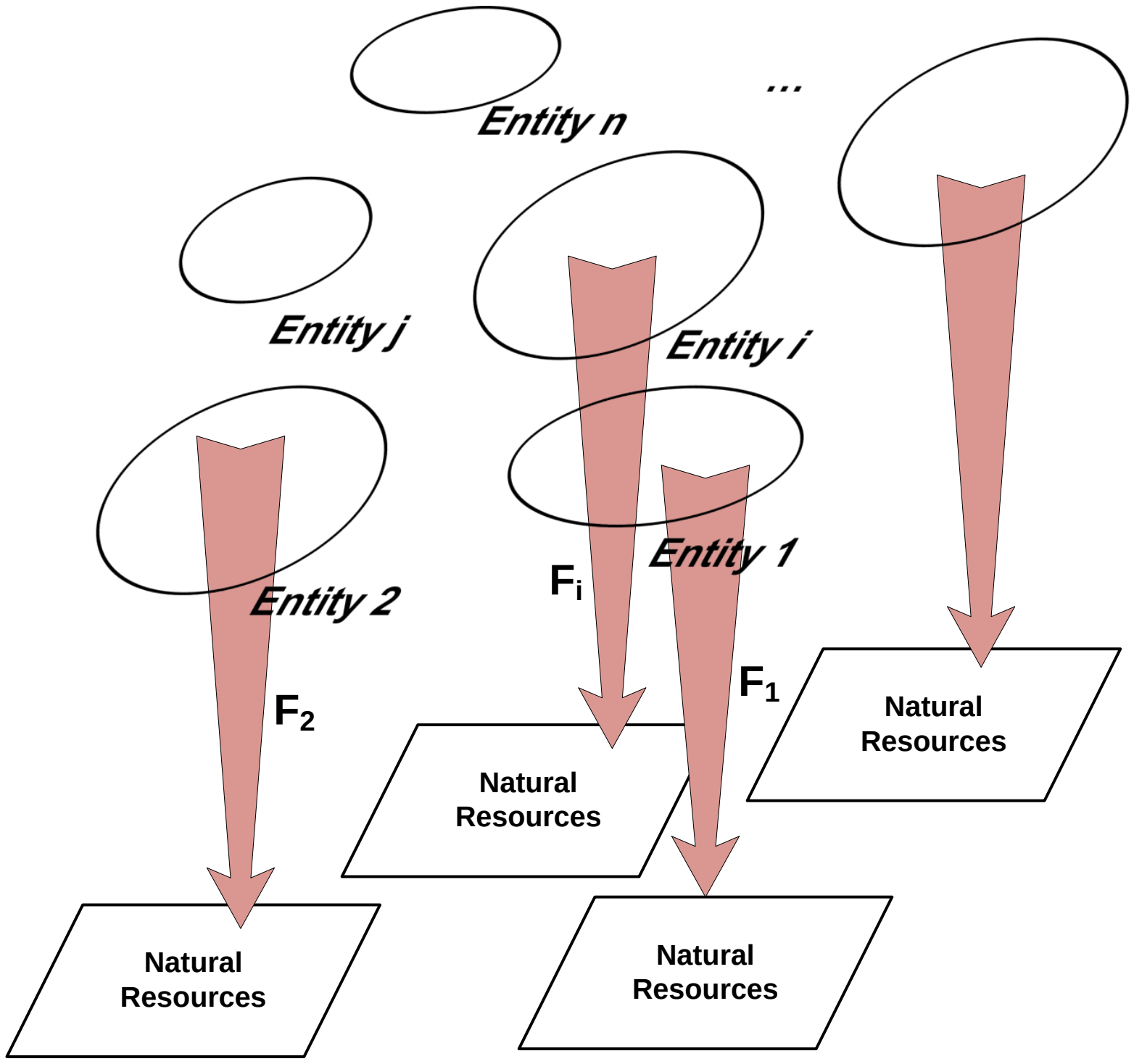

Figure 3: The *downward* actions of entities on the natural resources.

$f \in \mathcal{F}$:

$$\mathcal{F} = \{\text{Water}, \text{Energy}, \text{GHG}\}. \tag{2}$$

Also, $f$ will be assumed to be a numeric index later on for the purpose of simplifying the notation. In that case, $f = 1$ is associated with Water, $f = 2$ is associated with Energy, and $f = 3$ is associated with GHG Emissions.

If there is at least one action from an entity $j$ on another entity $i$, we set a directed edge between those entities. This leads us to a graph $G(V, E, w)$ among entities. To define the weight set $w$, the directed actions between every two entities are first aggregated on the *actor* entity and on the individual channel. To do so, the *maximum* price of each action's goods in a preselected reference market[5] is used to calculate the values of *inward* econons to a every *inactor* entity aggregated along their actors and channels. Let us denote the set for a typical entity $i$ and a typical channel $c$ as $W'_{i,c} = \left\{ w'_{i,c,j_{i,\theta}} \right\}$, where $\{j_{i,\theta}\}_\theta$ is the set of 'actor' entities on the entity $i$. As mentioned before, we do not recommend inter-channel normalization. Therefore, each channel is normalized separately, and the new sets of values is defined:

$$W_{i,c} = \left\{ w_{i,c,j_{i,\theta}} \right\} = \left\{ w'_{i,c,j_{i,\theta}} \Big/ \sum_{\theta'} w'_{i,c,j_{i,\theta'}} \right\}. \tag{3}$$

Using the sets $W_{i,c}$, we can now define the weights associated to the directed edges of the graph $G(V, E, w)$.

[5]In this work, we used the NYEC as the reference market.

Along with our discussion on importance of all channels regardless of the actual *hidden* value carried by them, we define the weight from an actor entity $j$ on an inactor entity $i$ by the maximum value from the subset $\{w_{i,c,j}\}_c$ defined as a subset of $W_{i,c}$ limited to only $j$ actions, i.e., $\underset{j_{i,\theta}=j}{\subset} W_{i,c}$:

$$w_{j \to i} = \max_c w_{i,c,j}. \tag{4}$$

Having the graph of the model, the set of responsibility among entities, $\mathbf{R} = \{R_{i,j}\}$, can be calculated according to the definition of *responsibility* in section 2. First, we need to calculate the unnormalized responsibility of the entity $j$ on the actions of the entity $i$, i.e., $\bar{R}_{j,i}$, which is the multiplications of the consequential immediate responsibility weights $w_{\cdot \to \cdot}$ along a path from $j$ to $i$ that *maximizes* that multiplication.

In order to calculate $R_{i,j}$ values on $G(V, E, w)$, we consider the shortest paths on its associated log-reversed graph $G^{\mathrm{LR}}(V, E, w^{\mathrm{LR}})$. As defined in section 2, $w^{\mathrm{LR}}_{j \to i} = \log\left(2\frac{w_{\max}}{w_{j \to i}}\right)$. The shortest path problem on directed graphs $G^{\mathrm{LR}}$ is a classic problem with various algorithms available to solve it. If we assume the shortest path from $j$ to $i$ on $G^{\mathrm{LR}}(V, E, w^{\mathrm{LR}})$ is the path $\beta^*$, then the unnormalized responsibility $\bar{R}_{i,j}$ can be calculated:

$$\bar{R}_{i,j} = \prod_{k=1}^{k_{\max}-1} w_{v_{\beta^*,k} \to v_{\beta^*,k+1}} = 2 \prod_{k=1}^{k_{\max}-1} \frac{w_{\max}}{\exp\left(w^{\mathrm{LR}}_{v_{\beta^*,k} \to v_{\beta^*,k+1}}\right)}. \tag{5}$$

By normalizing $\bar{R}$ on the actors side, we calculate the final responsibility $R$: $R_{i,j} = \bar{R}_{i,j} \Big/ \left(\sum_j \bar{R}_{i,j}\right)$.

The second part of the model describes how to *distribute* the immediate footprint of entities among them according to their responsibility $R$. The *actual* footprint of an entity $j$, denote $\mathbf{F}_j$ in the time interval $\Delta t$ is defined by using $R_{i,j}$ and $\bar{F}_{i,f}$:

$$\mathbf{F}_j = \{\mathbf{F}_{j,f}\}_f, \tag{6}$$

where

$$\mathbf{F}_{j,f} = \sum_{i=1}^{n} R_{i,j} \bar{F}_{i,f}, \tag{7}$$

is its *actual* footprint in the category $f$ of impacts on the nature, and $\bar{F}_{i,f}$ is the *immediate* footprint of the entity $i$ in the same category. In the next subsection, a hypothetical example is provided to show the properties of the proposed naive MEIO model.

### *3.1.1. An Illustrative Example of Four Fictional Entities*

In this section, a small example with four entities is presented. These entities, namely, A, B, C and D, form the set of EoI, and the rest of the world is assumed encapsulated in the entity RoW. The example shown in Figure 4 can be used to explain the details of how the framework works. Note that the interaction of Entity C with the rest of world is aggregated. In this example, a time interval (for example, a fiscal year) $[t_a, t_b]$ is considered in which 12 interactions happen. Each aggregated interaction is assumed to be associated with a time instance or a short time interval within $[t_a, t_b]$. In each channel and for each entity, the *inward* gauge econons are aggregated and combined considering the highest market price (in the reference market) of each item during the target time interval. For example, the Raw Material Channel of Entity C has two inward gauge econons: i) 2 ton Iron (at $t_7$) and ii) 1 ton Lead (at $t_8$). If we choose to use the NYEC as the reference market and the fiscal year of 2011 as the target interval, the value of these inward econons of Entity C will be \$358.52 (\$179.26 per ton) and \$2,681.02 respectively.[6] Also, it is worth mentioning again that although these econons are 'inward' to Entity C, Entity C is consider their 'actor.' In other words, the actor or source of an action receives the commodity, i.e., any *object* being transferred along the associated channel, of the associated econon, while the inactor or the sink of that action provides that commodity.

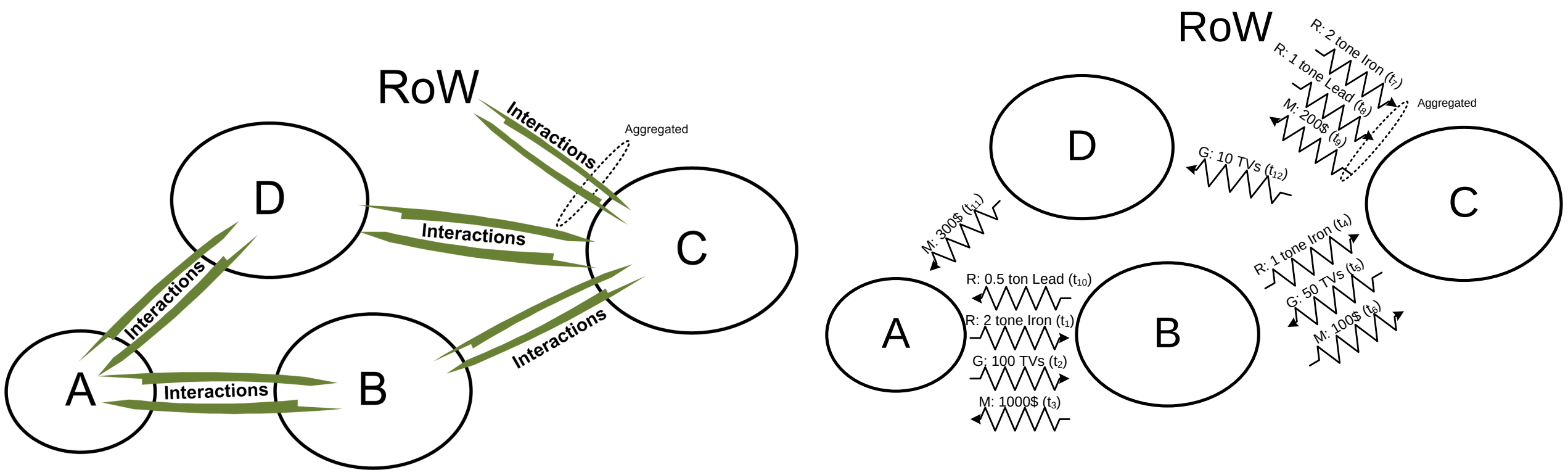

(a) (b)

Figure 4: An illustrative example on how the naive MEIO model works. a) The four entities that compose the EoI. b) The detailed gauge econons exchanged between the entities in a specific period of time.

For Entity B, although the 'outward' Raw Material Channel econons have different action time, they will be considered with the same price (the maximum price in the interval $[t_a, t_b]$): 1 ton Iron at the price of \$179.26 per ton and 0.5 ton Lead at the price of \$2,681.02 per ton. Therefore, Entities A and C have 88.20% and 11.80% shares in the Raw Material Channel of Entity B during the target interval. Table 2 summarized all channels of this example. The final step is the governing equations of the share-in-actions of the three entities. Briefly, total liability (footprint) of each entity is sum of its share in actions of each other entity times their actual liability (or footprint). The share-in-action ratio for each entity in actions of another entity is the maximum ratio among rations in each of four interaction channels. The list of gauge econons exchanged among the entities is also provided in Table 2. Also, in this table, the immediate responsibility of entities on footprint action of the others is presented in the percentage format. It is worth repeating the source-sink direction in the actions is in opposite direction of commodity transfer. For example, if 2 ton of Iron is transferred from Entity A to Entity B, the latter, i.e., Entity B, is the source and actor, and the entity A is the sink and inactor. Furthermore, the zero-valued $w_{i,j}$ are treated as N/A[7] from here on because there will be no actual edge in their place.

[6] For example: `http://www.indexmundi.com/commodities/?commodity=iron-ore&months=60&commodity=lead`.
[7] N/A: Not Available

| Sink Entity ($i$) | Channel | List of inward econons | Value of econons | $w_{i,j}$ | | | | |
|---|---|---|---|---|---|---|---|---|
| | | | | Source Entity ($j$) | | | | |
| | | | | A | B | C | D | RoW |
| A | 1 (Raw) | 2 ton Iron (B) | $358.52 (B) | N/A | 100% | 0% | 0% | 0% |
| A | 2 (Goods) | 100 TVs (B) | $20,000.00 (B) | N/A | 100% | 0% | 0% | 0% |
| A | 3 (Services) | N/A | N/A | N/A | N/A | N/A | N/A | N/A |
| A | 4 (Money) | N/A | N/A | N/A | N/A | N/A | N/A | N/A |
| A | All channels (Max) | — | $20,000.00 | 0% | 100% | 0% | 0% | 0% |
| B | 1 (Raw) | {1 ton Iron (C), 0.5 ton Lead (A) | {$179.26 (C), $1340.51 (A) | 88.20% | N/A | 11.80% | 0% | 0% |
| B | 2 (Goods) | N/A | N/A | N/A | N/A | N/A | N/A | N/A |
| B | 3 (Services) | N/A | N/A | N/A | N/A | N/A | N/A | N/A |
| B | 4 (Money) | {$1,000.00 (A), $100.00 (C) | {$1,000.00 (A), $100.00 (C) | 90.91% | N/A | 9.09% | 0% | 0% |
| B | All channels (Max) | — | $1,000.00 | 90.91% | 100% | 11.80% | 0% | 0% |
| C | 1 (Raw) | N/A | N/A | N/A | N/A | N/A | N/A | N/A |
| C | 2 (Goods) | {50 TVs (B), 10 TVs (D) | {$10,000.00 (B), $2,000.00 (D) | 0% | 83.33% | N/A | 16.67% | 0% |
| C | 3 (Services) | N/A | N/A | N/A | N/A | N/A | N/A | N/A |
| C | 4 (Money) | $200.00 (ROW) | $200.00 (ROW) | 0% | 0% | N/A | 0% | 100% |
| C | All channels (Max) | — | $10,000.00 | 0% | 83.33% | 100% | 16.67% | 0% |
| D | 1 (Raw) | N/A | N/A | N/A | N/A | N/A | N/A | N/A |
| D | 2 (Goods) | N/A | N/A | N/A | N/A | N/A | N/A | N/A |
| D | 3 (Services) | N/A | N/A | N/A | N/A | N/A | N/A | N/A |
| D | 4 (Money) | $300.00 (A) | $300.00 (A) | 100% | 0% | 0% | N/A | 0% |
| D | All channels (Max) | — | $300.00 | 100% | 0% | 0% | 100% | 0% |
| ROW | 1 (Raw) | {2 ton Iron (C), 1 ton Lead (C) | {$358.52 (C), $2,681.02 (C) | 0% | 0% | N/C | 0% | N/A |
| ROW | 2 (Goods) | N/A | N/A | N/A | N/A | N/A | N/A | N/A |
| ROW | 3 (Services) | N/A | N/A | N/A | N/A | N/A | N/A | N/A |
| ROW | 4 (Money) | N/A | N/A | N/A | N/A | N/A | N/A | N/A |
| ROW | All channels (Max) | — | N/C | 0% | 0% | N/C | 0% | 100% |

Table 2: The gauge econons involved in the illustrative example. The immediate responsibility $w_{i,j}$ is also provided.

| | A | B | C | D | RoW |
|---|---|---|---|---|---|
| A | 1.00 | 1.00 | 0.00 | 0.00 | 0.00 |
| B | 0.91 | 1.00 | 0.12 | 0.00 | 0.00 |
| C | 0.00 | 0.83 | 1.00 | 0.17 | 1.00 |
| D | 1.00 | 0.00 | 0.00 | 1.00 | 0.00 |
| RoW | 0.00 | 0.00 | 0.00 | 0.00 | 0.00 |

(a)

| | A | B | C | D | RoW |
|---|---|---|---|---|---|
| A | 1.00 | 1.00 | 0.12 | 0.02 | 0.00 |
| B | 0.91 | 1.00 | 0.12 | 0.02 | 0.00 |
| C | 0.76 | 0.83 | 1.00 | 0.17 | 0.00 |
| D | 1.00 | 1.00 | 0.12 | 1.00 | 0.00 |
| RoW | 0.00 | 0.00 | 0.00 | 0.00 | 1.00 |

(b)

| | A | B | C | D | RoW |
|---|---|---|---|---|---|
| A | 0.47 | 0.47 | 0.06 | 0.01 | 0.00 |
| B | 0.44 | 0.49 | 0.06 | 0.01 | 0.00 |
| C | 0.27 | 0.30 | 0.36 | 0.06 | 0.00 |
| D | 0.32 | 0.32 | 0.04 | 0.32 | 0.00 |
| RoW | 0.00 | 0.00 | 0.00 | 0.00 | 1.00 |

(c)

| | W | E | GHG |
|---|---|---|---|
| Unit | Lit | J | tCO2e |
| A | 10.00 | 100.00 | 1.00 |
| B | 1.00 | 10.00 | 0.10 |
| C | 20.00 | 20.00 | 1.50 |
| D | 5.00 | 7.00 | 3.50 |
| RoW | NA | NA | NA |

(d)

| | W | E | GHG |
|---|---|---|---|
| Unit | Lit | J | tCO2e |
| A | 12.22 | 58.96 | 2.05 |
| B | 12.81 | 59.95 | 2.09 |
| C | 8.05 | 13.61 | 0.74 |
| D | 2.91 | 4.47 | 1.22 |
| RoW | NA | NA | NA |

(e)

| | W | E | GHG |
|---|---|---|---|
| Unit | Lit | J | tCO2e |
| A | 4.6780 | 46.7799 | 0.4678 |
| B | 0.4886 | 4.8857 | 0.0489 |
| C | 6.0438 | 6.0438 | 0.4533 |
| D | 1.6036 | 2.2450 | 1.1225 |
| RoW | NA | NA | NA |
| Total | 12.8139 | 59.9544 | 2.0925 |
| Change | 12.8139 | 5.9954 | 20.9245 |

(f)

Table 3: The naive MEIO model results for the illustrative example. a) The weights $w$, b) The unnormalized responsibility $\bar{R}$, c) The responsibility $R$, d) The immediate footprint, e) The actual footprint, and f) The breakdown of the actual footprint of Honda based on its sources. Please note the change values are not in percentage.

The immediate responsibility weights are represented by $w$, which is expected to be sparse because many of actors do not have direct interactions with each other. In contrast, The final responsibility is represented by $\mathbf{R}$ would be a highly dense. In Table 3, the $w$, $\bar{R}$, $R$, and also actual footprint $\bar{F}$, final footprint $F$, and ultimately the breakdown of the final footprint of one of entities are presented. It is assumed that the actual footprint of Entities A, B, C, and D are (10 Liter, 100 J, 1 $tCO_2e$), (1 Liter, 10 J, 0.1 $tCO_2e$), (20 Liter, 20 J, 1.5 $tCO_2e$), and (5 Liter, 7 J, 3.5 $tCO_2e$), respectively (as provided in Table 3(d)). It is interesting that Entity B is charged with much higher footprint compared to its actual footprint that shows how the responsibility of the entities enables the MEIO model to allocate the footprint to those who are responsible. The breakdown of the final footprint changed to Entity B, Table 3(f), gives the detailed information on how the allocation is distributed among the entities whom Entity B is a direct or indirect actor on. This information could be used and leveraged by the decision maker of Entity B to seek changes in behavior, interactions, and business strategies to lower total or specific portion of its footprint. As discussed in Farrahi Moghaddam et al. (2014), addressing the total and aggregated footprint would result in missing considerable opportunities to adapt decisions that could highly impact a specific subset of entities or natures. Although such actions could have small or even negligible effect on the total footprint Entity B, in this example, they could highly noticeable for the other entities especially in terms of development. For example, some types of air pollution do not extend beyond some local regions, and therefore those regions should be considered as *separate* entities in the model. We think to address this paradox it is required to develop a new 'distance' to calculate the impact of a sustainability project. This distance function, which would not focus on the aggregated footprint, will be the focus of another work.

The final and actual footprint relations calculated using the MEIO model can be also written in the form of a set of equations:

$$F_{A,f} = 0.47\bar{F}_{A,f} + 0.44\bar{F}_{B,f} + 0.27\bar{F}_{C,f} + 0.32\bar{F}_{D,f}, \tag{8}$$

$$F_{B,f} = 0.47\bar{F}_{A,f} + 0.49\bar{F}_{B,f} + 0.30\bar{F}_{C,f} + 0.32\bar{F}_{D,f}, \tag{9}$$

$$F_{C,f} = 0.06\bar{F}_{B,f} + 0.06\bar{F}_{B,f} + 0.36\bar{F}_{C,f} + 0.04\bar{F}_{D,f}, \tag{10}$$

$$F_{D,f} = 0.01\bar{F}_{B,f} + 0.01\bar{F}_{B,f} + 0.06\bar{F}_{C,f} + 0.32\bar{F}_{D,f}. \tag{11}$$

It is worth repeating again that the footprint relations are independent from the footprint categories $f$.

It can be observed from the illustrative example that there could be a big difference between the immediate and final footprint of an entity when an MEIO model is used. However, in contrast to scope-based approaches, where the big entities account for the footprint of their *dependent* smaller entities, the final footprint calculated using the MEIO model does not assume any restriction or dependability among the entities. Moreover, it does not require a known economic relation among the entities, which makes it distinct from the MRIO models. This can be seen mainly because of the MEIO model's approach of considering actions themselves as 'objects.' This way of modeling, which is inspired from the particle physics, will be more discussed in section 3.1.3 and also Appendix B.

*3.1.2. Sensitivity analysis of the strongest path*

It is worth noting that the results of the MEIO models could suffer from uncertainty in the input data. In order to have a measure of the degree of variability and uncertainty of the results, we introduce a sensitivity analysis following that has been considered for the shortest path problem (Bilò et al., 2010; Buhmann et al., 2013; Šrámek, 2013).

In the MEIO models, the unnormalized responsibility of a node $j$ on another node $i$ is obtained along the strongest path(s) from $j$ to $i$. Any noise or error in the value of the gauge econons would affect the selection of the strongest path. In order to have an estimation of the possible variability of the result, we consider the stability number of the strongest path: $\delta\left(S_{i,j}\right) = \sigma_{S_{i,j}}$. This means, with $\lambda = \delta\left(S_{i,j}\right)$, there is a $\bar{\lambda}$ perturbation that requires a strongest path different from $S_{i,j}$ at least on one edges. Therefore, the second strongest path on the original graph is at least $\delta\left(S_{i,j}\right)$ weaker than $S_{i,j}$. Thus, we can introduce a minimum variability of $v_{i,j} = \left(1 + \delta\left(S_{i,j}\right)\right)/2$. Here, we consider a smoothing in the form of an average with the extreme case of having the second strongest path as strong as the first one. Therefore, we define the *variation interval* of unnormalized responsibility to be:

$$\left[\bar{R}_{i,j}\left(1 + \sigma_{S_{i,j}}\right)/2, \bar{R}_{i,j}\left(3 - \sigma_{S_{i,j}}\right)/2\right].$$

For instance, in our illustrative example, if we consider $j = \mathrm{A}$ and $i = \mathrm{C}$, the strongest path is $\beta_1 = \{\mathrm{A}, \mathrm{B}, \mathrm{C}\}$ with an associated unnormalized responsibility of 0.76 (see Table 3(b)). The second strongest path from A to C would be along D with an associated (unnormalized) strength of 0.17: $\beta_2 = \{\mathrm{A}, \mathrm{D}, \mathrm{C}\}$. Therefore, the stability number would be $\delta\left(S_{\mathrm{C,A}}\right) = \sqrt{(0.17/0.76)} = 0.47$. Here, we use the square root because the path $\beta_1$ has a length of 2 edges. Our definition of variability gives $v_{\mathrm{C,A}} = (1 + 0.47)/2 = 0.74$, which results in an variation interval of unnormalized responsibility from of A on C equal to $[0.76\,(1+0.47)\,/2, 0.76\,(3-0.47)\,/2] = [0.58, 0.96]$. In addition, it is worth comparing this value with an *exact* calculation made directly using the second strongest path data: $[(0.76+0.17)\,/2, 0.76+(0.76-(0.76+0.17)\,/2)] = [0.47, 1.06]$. In a future work, where we will consider examples with multiple paths among the vertexes, we will discuss the importance and role of variability interval in details.

### *3.1.3. A Note on the difference between the MEIO and many MRIO models*

As mentioned before, an explicit deviation between the MEIO and the MRIO models is relaxation of the regional dependency requirement in the MEIO case. In the MEIO models, an actor could belong to one region, to multiple regions, or even to *no* region. However, the main difference between these two approaches to modeling is in their perspective with respect to the behavior. The MEIO models do not require a "model" of an entity's behavior. Instead, they directly work with its actual actions. In addition, the MRIO models are usually quasi-static because of many constraints involved and also the requirement to limit the uncertainty. In contrast, an MEIO model uses a *responsibility* approach that does not require full understanding of the mechanisms *driving* the actions and the interactions. This is of interest because, in our entangled global world, interactions are highly multi optional and decision dependent. However, this also means that the current form of the MEIO modeling is unable to provide insights and forecast on actions that may be placed in the future (we will discuss this concern in the Appendix B). In this work, the proposed MEIO models are used to estimate the responsibility of an entity for actions of a time interval in the 'past.'

## *3.2. First Main Model: Provider-Perspective (P-Perspective) MEIO Model*

Application of the naive MEIO model to the *whole* world, which would include all providers and end users of all sectors, is theoretically feasible. However, there would be some practical challenges, especially with respect of data collection and privacy at the end users side. Because our main target here is to estimate and calculate the footprint and responsibility of a provider(s) and also an end user in a specific sector, we will introduce two variations of the MEIO model that are adapted to these 'perspectives.' The first variation, the Provider-perspective (P-perspective) MEIO model, focuses on calculating the final footprint of a provider while preventing footprint leakage to or from the end users side.

In this model, the modified world $V$ comprised of an ecosystem of the End Users or Ultimate Users (UUs) entity, the RoW entity, and set of EoI (Figure 5(a)). The EoI consists of the provider of interest and also all other entities that directly or indirectly interact with that provider, such as utilities who provide low level infrastructure services, such as electricity, water, and broadband. Examples of the provider of interest could be Auto makers or Telco enterprises. It is worth noting the model can handle more than one provider of interest. However, in this work, we focus on only one-provider cases. It is possible that some of those entities, such as electricity utilities, also have direct interactions with the end users. However, in the P-perspective MEIO model, we ignore such interactions. To formalize this sector-limited modeling, we introduce the sector *slicing hyper-line*. As mentioned before, this hyper-line limits the visibility of or to the end users only to those interactions that are associated to the sector of the provider. The concept of slicing hyper-line is shown in Figure 5(b) separating the UUs and the EoI. More specific partitioning of the entities based on their function, location, among other factors is possible.[8] However, in the proposed model in this work we aggregate all UUs in one entity, and all other factors except the interactions are ignored. Similar to the naive model, every entity has its own immediate footprint on the natural resources (see Figure 5(d)). The slicing hyper-line enforces that only that footprint of the UUs associated to the product or service they received from the provider is visible in the P-perspective and vice versa. Also, because the UUs act as the sink of econons on the money channel to the provider, the aggregated interaction between UUs and RoW on this channel is also included in the model. After imposing these restrictions, the rest of the P-perspective MEIO model is the same as the naive model. The edge weights on every channel and then for all channels can be calculated, as described in section 3.1, and be then used to calculate the unnormalized and final responsibilities. In sections 4.1 and 4.2, two use cases will be provided to show how this model works in practice. However, it is worth highlighting some of its specific features here. First, the provider is not required to take responsibility of the whole footprint associated to its end users' utilization of its *product* or *service*. This makes the MEIO models a new alternative to the scope-based assessments. Second, the upstream and downstream interactions and also their associated footprint are directly included in the model without requiring any special treatment of such interactions. In other words, the MEIO model can intrinsically cover many features of LCA with proper selection of analysis time interval and possibly a hierarchical generalization with respect to time intervals.

### *3.3. Second Main Model: End User-Perspective (E-Perspective) Model*

The last model proposed in this work is the End user-perspective (E-perspective) MEIO model. The need for this model can be justified form the behavior aspect of the footprint and impact assessment. Behavioral changes, in the form of increasing sustainable behaviors, decreasing unsustainable behaviors, and replacing behaviors toward sustainability (see Appendix A for more details), are important aspect of any sustainability solution. In order to encourage or promote behavioral changes in the end users, it is important to provide a 'personalized' footprint assessment for every end users. Although *adjusted* share of an end users from the total final footprint of the UUs entity in the P-perspective MEIO model seems to be a good candidate for such an assessment, it could be ineffective because of the curse of big numbers; the share of an individual end users could become so small and negligible that it could not initiate any behavioral change. This is in contrast to the general understanding of economic and business drive of many products and services in which the absence of the end users would collapse the whole sector. This shows that every end user has a considerable *responsibility* in actions of the providers and also other entities. In order to calculate such a personal responsibility, we propose the E-perspective MEIO model in this section.

The main difference between the P-perspective and E-perspective MEIO models is that the UUs entity is merged with the RoW entity in the E-perspective. Only one particular end user, say End User A, is kept outside the RoW in order to calculate its associated personal responsibility and also footprint. This means that the modified world $V$ comprised of an ecosystem of the End User A, the RoW, and the set of EoI. Compared to the P-perspective model, the only weights that would be updated are those of edges connecting the End User A to the provider because the edges between A and other entities in the EoI are filter out by the sector slicing hyper-line (see Figure 6). The rest of calculation is the same as for the naive and P-perspective models, by starting to calculate the directed graph weights $w$, the unnormalized responsibilities $\bar{R}$, and the final responsibilities $R$ that can be used in combination of immediate footprint $\bar{F}$ to finally calculate the final footprint $F$. In the use cases, it will be seen that the footprint of the End User A in the E-perspective modeling would be considerably and *correctly* higher than its simple dividend share of the UUs in the P-perspective modeling of the same use case. Also, it is worth mentioning that although we will use a 'typical' end user in the use cases, the assessments can be personalized to a specific end user in actual use cases taking into account their actual interactions and footprint.

[8] Even, representations in higher dimensions, such as hyper-cones of Figure 5(c) could be foreseen.

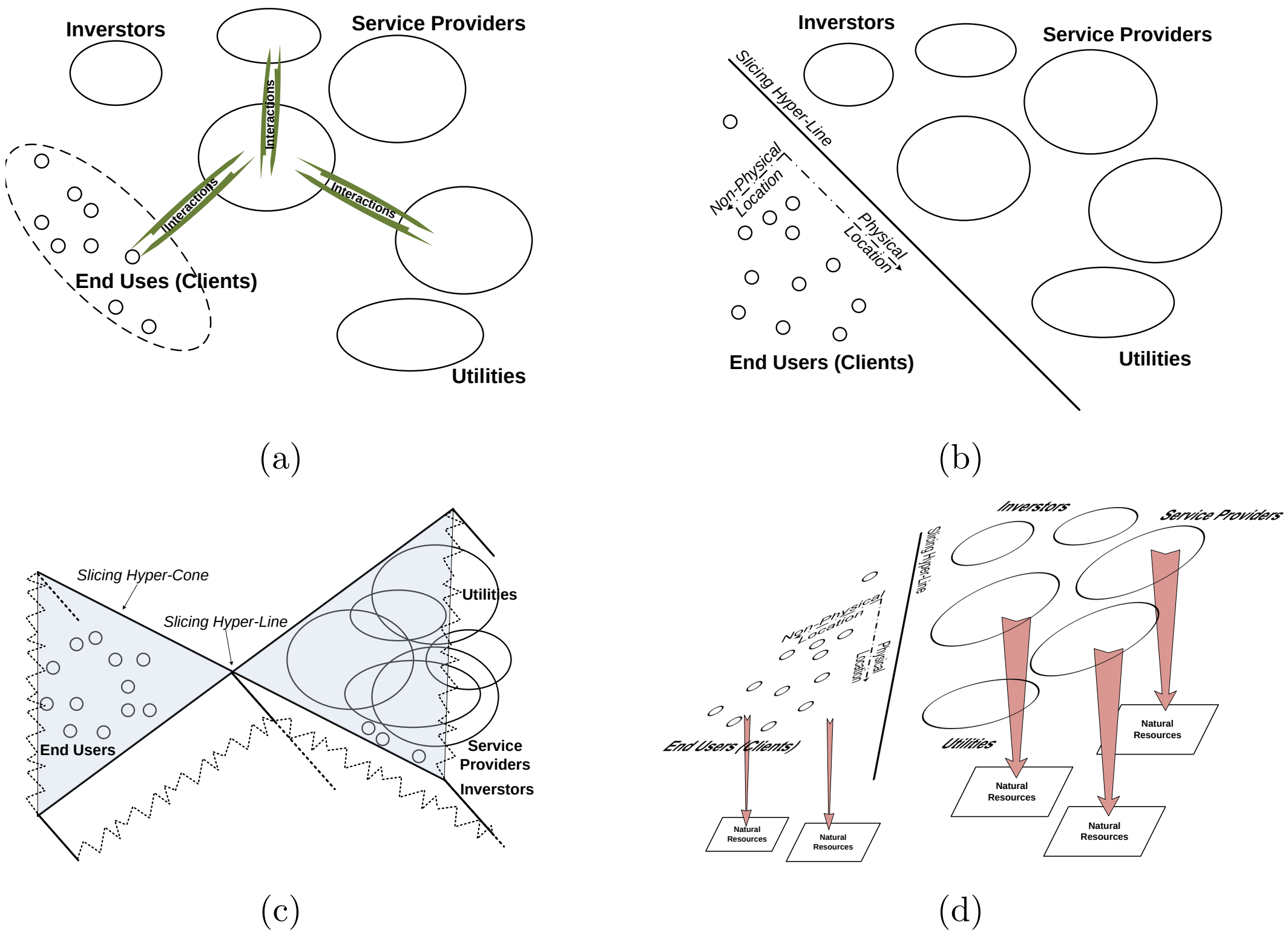

Figure 5: The MEIO framework. a) End users and the EoI in the P-perspective MEIO model. b) The sector slicing hyper-line separate the end users from the other entities. c) The entities could be imagined in high-dimension representation. d) The entities' action on the environment are modeled as downward actions on natural resources.

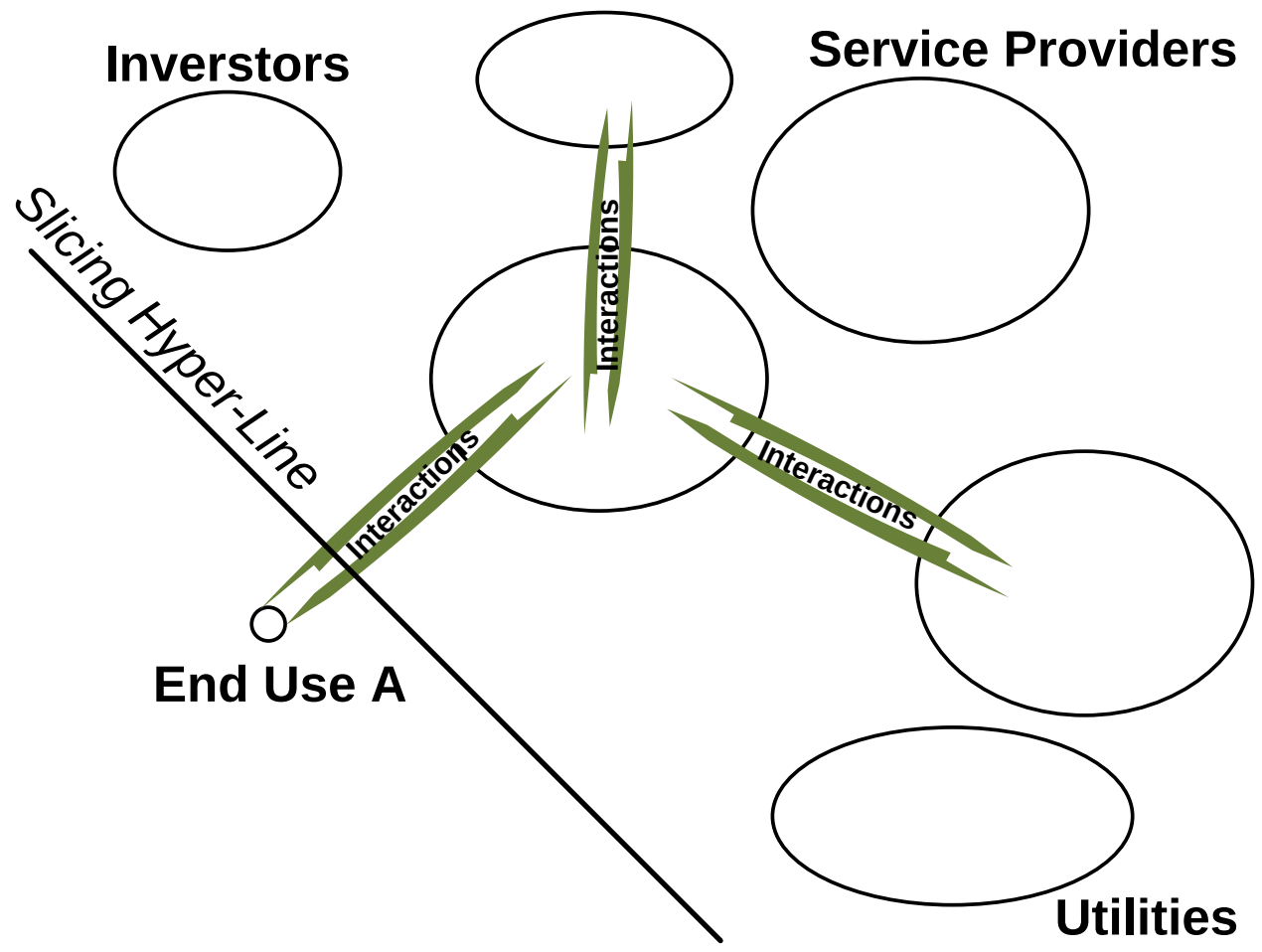

Figure 6: The E-perspective MEIO model.

### *3.4. Real-time and 'single' action assessments*

The object-oriented and action-based nature of the MEIO modeling makes it unnecessary to have the model's variables as [continuous] functions along time. The sequence of actions by itself determines the interactions and the responsibilities. This allows us to easy perform finer-granular analyses: i) for just a single action, ii) for a set of 'exclusive' actions that share a common short interval of time, and even iii) for a set of actions with no 'exclusive' shared interval of time. This perspective could be very important especially considering that an end user's actions are more similar to a set of spontaneous events rather than 'flow'-alike variables. We will develop this aspect of the MEIO models in the future.

## 4. Experimental results

In this section, two use cases from Auto and Telco industries are provided as *illustrative* examples on how the P-perspective and E-perspective MEIO models could be used in actual product/service providing settings. As mentioned before, these use cases are not *eligible* for being cited for sustainability performance evaluation of the providers involved. Only the public, and possibly incomplete, data available to the authors were used in the calculations, and we avoided any data request to the entities involved in these use cases in order to prevent any implicit external influence on the models. We will modify and update the results based on a more complete data that could be provided to us by these entities in the future. However, for the main purpose of showing *how* the responsibility and footprint should be *allocated* among the entities, which is a task independent from the degree of accuracy of the data used, these use cases can serve as illustrative examples.

### *4.1. Honda Use case*

Honda Motor Co. Ltd was the first Auto maker company who disclosed its GHG emissions in a Scope-3 analysis (Honda Motor Co. Ltd., 2012). This motivated us to consider them in our use case for the Auto makers industry. Although the reported Honda's Scope-3 footprint were for FY2011, we hypothetically use it for FY2012 and then transfer it back to FY2011 in order to add *noise* and confusion in the data to make it incomparable with actual footprint. This is along with our intention that we are not looking to assess a provider but instead we are studying how to allocate its footprint. In this way, we calculate an imaginary footprint of 234.4 Mt$CO_2$e associated to the FY2011 buyers compared to that of 195.9 Mt$CO_2$e reported in Honda Motor Co. Ltd. (2012). For more details, please see Supplementary Material S.1.

| | Natural Gas | Liquefied Petroleum Gas | Diesel | Total Footprint |
|---|---|---|---|---|
| GHG Factor (Kt$CO_2$e/GLit) | 1.89 | 2,734.40 | 2,785.50 | |
| Water Factor (Gal/kWh) | 0.20 | 0.72 | 0.72 | |
| Conversion Factor (GJ/MLit) | 39.00 | 25,700.00 | 39,600.00 | |
| Honda Energy Consumption (GJ) | 8,112,726.00 | 3,426,524.00 | 1,334,611.00 | |
| Honda Energy Footprint (GWh) | 2,253.72 | 951.89 | 370.75 | 3,576.36 |
| Honda Energy-related GHG Footprint (Scope 1) (Kt$CO_2$e) | 393.16 | 364.57 | 93.88 | ~~851.61~~ |
| Honda GHG Footprint (Scope 1) (Kt$CO_2$e) | | | | 1,240.00 |
| Honda Energy-related Water Footprint (Scope 1) (KGal) | 450.74 | 685.36 | 266.94 | ~~1,403.04~~ |
| Honda Water Footprint Tap (KGal) | | | | 4,165.20 |
| Honda Water Footprint Ground (KGal) | | | | 3,466.20 |
| Honda Water Footprint Total (KGal) | | | | 7,631.40 |

Table 4: The *immediate* Honda's footprint. The water consumption for on-site electricity generation is not summed because we were not sure it has been already accounted for. The Liter unit in the footprint factors refer to the fuel used.

In this use case, we consider the FY2011 as the $\Delta t$. Also, the set of EoI is considered to be: **Honda** The Honda Motor Co. Ltd, **Investors** The Honda's investors, **MatSteel** The Steel industry across the globe, **MatAl** The Aluminum industry across the globe, **UtlJP** The electricity utilities in Japan, **UtlNA** The electricity utilities in North America, **UtlSA** The electricity utilities in South America, **UtlEU** The electricity utilities in European Union, **UtlAO** The electricity utilities in Asia-Oceanic, and **UtlCH** The electricity utilities in China. As mentioned before, the set of EoI could be incomplete and many of them have been aggregated across big geographical regions. However, for the purpose of our illustration, the EoI covers main energy and material actors in addition to the main provider, and especially its investors. The end users, i.e., who bought a car or motorcycle in the FY2011, are represented by the **Buyers** entity. The details on how the footprint of each entity is calculated and also on how the econons exchanged among them are evaluated along the data sources are provided in Supplementary Material S.1. Here, the footprint of the main provider is provided in Table 4. For the water and GHG footprint, the footprint calculated based on the reported energy consumption using different types of fuels is ignored, and only the reported footprint by the provider is considered in order to avoid double counting. Finally, the immediate footprint of all entities is provided in Table 5.

Below, in two cases, we estimate the share of the Auto maker and also the share of an individual car buyer using the MEIO models. However, it is worth mentioning that the provider, and probably the end user, has performed many nature-friendly actions that have not been accounted for in this illustrative analysis, and a few of them are listed below to acknowledge them for future studies: • Reducing the electricity demand on peak hours to avoid non-green sources, • Recycle, reuse, and waste reduction initiatives, and • Contribution to local communities.

#### *4.1.1. Honda's footprint in 2011 (P-perspective)*

In this section, the P-perspective MEIO model is applied to the FY2011 of Honda. The full data is described in Supplementary Material S.1. The final gauge econons transferred among the entities is provided in Table 6, where the weight $w$ of the associated MEIO graph are also provided in a percentage scale. The weight $w$, which is also provided in the form of a matrix in Table 7(a), can be used to calculate the unnormalized responsibility and then final responsibility among the entities as described in section 3.2. The final responsibility is provided is Table 7(b), and can be combined with the immediate footprint of the entities, Table 7(c), to calculate the final footprint of every entity, including that of the product provider under study, which is shown in Table 7(d).

| Actor | Water Footprint (Mgal) | Energy Footprint (GWh) | GHG Footprint ($MtCO_2e$) |
|---|---|---|---|
| Honda | 7.63 | 3,576.36 | 1.24 |
| Buyers | 0.00 | 972,950.00 | 234.40 |
| Investors | 0.00 | 0.00 | 0.00 |
| MatSteel | 27,000,000.00 | 8,290,500.00 | 10,530.78 |
| MatAl | 366,470.00 | 1,930,610.00 | 1,996.46 |
| UtlJP | 508,670.00 | 997,400.00 | 417.51 |
| UtlNA | 2,209,520.00 | 4,701,070.00 | 2,983.31 |
| UtlSA | 280,640.00 | 584,660.00 | 82.43 |
| UtlEU | 1,490,600.00 | 3,635,600.00 | 1,413.08 |
| UtlAO | 511,630.00 | 1,346,420.00 | 1,030.95 |
| UtlCH | 1,961,900.00 | 3,503,400.00 | 2,959.85 |

Table 5: The immediate footprint of various actors in the Honda P-perspective use case.

| Source | Channel | List of econons | Value of econons | Non-zero Elements of $w$ |
|---|---|---|---|---|
| Honda | 2 (Goods) | 22,185,000 vehicle (Buyers) | $M107,479 (Buyers) | $w$(Honda, Buyers) = 100% |
| Honda | 4 (Money) | $M6,423 (Investors),<br>$M217.31 (MatSteel),<br>$M76.13 (MatAl),<br>$M247.04 (UtlJP),<br>$M107.49 (UtlNA),<br>$M32.23 (UtlSA),<br>$M15.84 (UtlEU),<br>$M84.99 (UtlAO),<br>$M43.50 (UtlCH),<br>$M100,231.5 (RoW) | $M6,423 (Investors),<br>$M217.31 (MatSteel),<br>$M76.13 (MatAl),<br>$M247.04 (UtlJP),<br>$M107.49 (UtlNA),<br>$M32.23 (UtlSA),<br>$M15.84 (UtlEU),<br>$M84.99 (UtlAO),<br>$M43.50 (UtlCH),<br>$M100,231.5 (RoW) | $w$(Honda, Investors) = 5.98%,<br>$w$(Honda, MatSteel) = 0.20%,<br>$w$(Honda, MatAl) = 0.07%,<br>$w$(Honda, UtlJP) = 0.23%,<br>$w$(Honda, UtlNA) = 0.1%,<br>$w$(Honda, UtlSA) = 0.03%,<br>$w$(Honda, UtlEU) = 0.01%,<br>$w$(Honda, UtlAO) = 0.08%,<br>$w$(Honda, UtlCH) = 0.04% |
| Buyers | 4 (Money) | $M107,479 (Honda) | $M107,479 (Honda) | $w$(Buyers, Honda) = 100% |
| Investors | All | — | — | $w$(Buyers, Honda) = 100% |
| MatSteel | 1 (Raw) | 241,459.50 t Steel (Honda),<br>1,499.76 Mt Steel (ROW) | $M217.31 (Honda),<br>$B1,349.78 (ROW) | $w$(MatSteel, Honda) = 0.02%,<br>$w$(MatSteel, ROW) = 99.98% |
| MatSteel | 4 (Money) | $M494.45 (UtlAO) | $M494.45 (UtlAO) | $w$(MatSteel, UtlAO) = 100% |
| MatAl | 1 (Raw) | 28,540.50 t Steel (Honda),<br>44.07 Mt Steel (ROW) | $M76.13 (Honda),<br>$M117,557.09 (ROW) | $w$(MatAl, Honda) = 0.07%,<br>$w$(MatAl, ROW) = 99.93% |
| MatAl | 4 (Money) | $M3,916.38 (UtlAO) | $M3,916.38 (UtlAO) | $w$(MatAl, UtlAO) = 100% |
| UtlJP | 1 (Raw) | 1.60 TWh Elect (Honda),<br>995.80 TWh Elect (ROW) | $M247.04 (Honda),<br>$M153,750 (ROW) | $w$(UtlJP, Honda) = 0.16%,<br>$w$(UtlJP, ROW) = 99.84% |
| UtlNA | 1 (Raw) | 1.54 TWh Elect (Honda),<br>4,699.53 TWh Elect (ROW) | $M107.49 (Honda),<br>$M328,030 (ROW) | $w$(UtlNA, Honda) = 0.03%,<br>$w$(UtlNA, ROW) = 99.97% |
| UtlSA | 1 (Raw) | 0.33 TWh Elect (Honda),<br>584.33 TWh Elect (ROW) | $M32.23 (Honda),<br>$M57,069.56 (ROW) | $w$(UtlSA, Honda) = 0.06%,<br>$w$(UtlSA, ROW) = 99.94% |
| UtlEU | 1 (Raw) | 0.15 TWh Elect (Honda),<br>3,635.45 TWh Elect (ROW) | $M15.84 (Honda),<br>$M383,903.5 (ROW) | $w$(UtlEU, Honda) = 0.004%,<br>$w$(UtlEU, ROW) = 99.996% |
| UtlAO | 1 (Raw) | 0.95 TWh Elect (Honda),<br>1,345.47 TWh Elect (ROW) | $M84.99 (Honda),<br>$M120,370 (ROW) | $w$(UtlAO, Honda) = 0.04%,<br>$w$(UtlAO, ROW) = 99.96% |
| UtlCH | 1 (Raw) | 0.58 TWh Elect (Honda),<br>3,502.82 TWh Elect (ROW) | $M43.50 (Honda),<br>$M262,711.5 (ROW) | $w$(UtlCH, Honda) = 0.02%,<br>$w$(UtlCH, ROW) = 99.98% |

Table 6: The gauge econons and also related responsibility graph's weights in the Honda P-perspective use case.

It is interesting to point out that the P-perspective Honda's GHG footprint is estimated at 120.15 $MtCO_2e$ using the MEIO model, and it is almost half of the Buyers' GHG footprint, i.e., 234.4 $MtCO_2e$. This shows the advantage of the MEIO modeling compared to scoping approaches because the MEIO models enable allocation of footprint based on the responsibility of actors without overcharging them with the footprint of other actor who may not totally stand within the territory of the actor under study.

A second interesting point can be seen from Table 7(e). In this table, the breakdown of the footprint of the provider based on the other entities is presented. This analysis, which is simply the $R_{i,\text{Honda}}\bar{F}_i$, enables the provider's decision makers to focus on the specific interactions and econons in order to reduce the total or a portion of their footprint. In addition, it can be seen that a big portion of the Honda's final footprint is sourced from material providers, especially in terms of water footprint, and this shows that the MEIO models implicitly provide a life cycle analysis in terms of upstream, downstream, and sidestream interactions, as defined in the flatten LCA model (Farrahi Moghaddam et al., 2014), which makes the MEIO modeling a possible framework to encompass the LCA approaches. The implicit share of the provider in the footprint of the other entities, such as that of the Steel industry, would eventually guide the provider toward initiatives that indirectly help those industries to reduce its footprint (Wörtler et al., 2013). Finally, it is worth noting that the final footprint of the provider is several order of magnitude bigger than its *immediate* footprint. For example, the Honda's GHG footprint is 96.90 times bigger than its immediate Scope-1 GHG footprint of 1.24 $MtCO_2e$. This confirms the general consensus that generalized analyses, such as Scope-3 assessment, LCA, and in particular the proposed MEIO modeling, are required in order to estimate a more accurate footprint of any entity, especially the providers.

Another interesting result is the final footprint associated to the investors. Although this entity was not assigned with any immediate footprint, it is responsible for non-negligible final footprint. For example, its GHG footprint is estimated at 7.21 $MtCO_2e$. This shows that the MEIO models are capable to propagate the responsibilities to farthest entities without any requirement for iteration or spacial consideration if the proper gauge econons are considered in the data. In addition, it emphasizes on the fact that the investors have a more important role than of risk management of their investments, and they should take responsibility in the actions and footprint of their investees.

### *4.1.2. A buyer's footprint*

In this section, we repeat the calculation of previous section this time in the End User perspective. The main changes are merging the Buyers entity with the RoW and adding a **BuyerA** entity that represents an individual car buyer in FY2011. Using the data provided in Supplementary Material S.1, we arrive to the table of gauge econons presented in Table 8. It is worth observing that BuyerA, who is a typical average buyer in this use case, has a very small weight of $w_{\text{BuyerA}\to\text{Honda}} = 1.86 \times 10^{-5}$. However, as will be seen in the final results, this small weight will result in a considerable footprint for BuyerA.

Repeating the same procedure of the E-perspective MEIO modeling to calculate the unnormalized responsibility $\bar{R}$, the final responsibility $R$, and the final footprint $F$, we obtain the results provided in Table 9. As said before, the E-perspective MEIO models can be only used to estimate the footprint of an individual end user, and their results for the other entities are not accountable because a big portion of actors, i.e., the rest of the end users, are moved to the RoW and are shadowed.

If we look at the footprint breakdown of the BuyerA footprint in Table 9(e), its main GHG footprint comes from the MatSteel and MatAl entities. This is in contrast with the general understanding that a car by itself is a major immediate GHG emitters. This also shows that these material industries have a high ratio of GHG to energy consumption, and therefore is a lot of opportunities to improve their performance. The responsibility-based GHG footprint of the BuyerA entity is estimated to be 8.38 times bigger than its immediate GHG emissions of 18.21 $KgCO_2e$. Furthermore, it can be observed that for another footprint category, for example energy consumption, the BuyerA's self footprint is the main

(a)

| | Honda | Buyers | Investors | MatSteel | MatAl | UtlJP | UtlNA | UtlSA | UtlEU | UtlAO | UtlCH | RoW |
|---|---|---|---|---|---|---|---|---|---|---|---|---|
| Honda | 1.00 | 0.94 | 0.06 | 0.00 | 0.00 | 0.00 | 0.00 | 0.00 | 0.00 | 0.00 | 0.00 | 0.00 |
| Buyers | 1.00 | 1.00 | 0.00 | 0.00 | 0.00 | 0.00 | 0.00 | 0.00 | 0.00 | 0.00 | 0.00 | 0.00 |
| Investors | 1.00 | 0.00 | 1.00 | 0.00 | 0.00 | 0.00 | 0.00 | 0.00 | 0.00 | 0.00 | 0.00 | 0.00 |
| MatSteel | 0.00 | 0.00 | 0.00 | 1.00 | 0.00 | 0.00 | 0.00 | 0.00 | 0.00 | 0.00 | 0.00 | 0.00 |
| MatAl | 0.00 | 0.00 | 0.00 | 0.00 | 1.00 | 0.00 | 0.00 | 0.00 | 0.00 | 0.00 | 0.00 | 0.00 |
| UtlJP | 0.00 | 0.00 | 0.00 | 0.00 | 0.00 | 1.00 | 0.00 | 0.00 | 0.00 | 0.00 | 0.00 | 0.00 |
| UtlNA | 0.00 | 0.00 | 0.00 | 0.00 | 0.00 | 0.00 | 1.00 | 0.00 | 0.00 | 0.00 | 0.00 | 0.00 |
| UtlSA | 0.00 | 0.00 | 0.00 | 0.00 | 0.00 | 0.00 | 0.00 | 1.00 | 0.00 | 0.00 | 0.00 | 0.00 |
| UtlEU | 0.00 | 0.00 | 0.00 | 0.00 | 0.00 | 0.00 | 0.00 | 0.00 | 1.00 | 0.00 | 0.00 | 0.00 |
| UtlAO | 0.00 | 0.00 | 0.00 | 0.00 | 0.00 | 0.00 | 0.00 | 0.00 | 0.00 | 1.00 | 0.00 | 0.00 |
| UtlCH | 0.00 | 0.00 | 0.00 | 0.00 | 0.00 | 0.00 | 0.00 | 0.00 | 0.00 | 0.00 | 1.00 | 0.00 |
| RoW | 0.00 | 0.00 | 0.00 | 0.00 | 0.00 | 0.00 | 0.00 | 0.00 | 0.00 | 0.00 | 0.00 | 0.00 |

(b)

| | Honda | Buyers | Investors | MatSteel | MatAl | UtlJP | UtlNA | UtlSA | UtlEU | UtlAO | UtlCH | RoW |
|---|---|---|---|---|---|---|---|---|---|---|---|---|
| Honda | 0.50 | 0.47 | 0.03 | 0.00 | 0.00 | 0.00 | 0.00 | 0.00 | 0.00 | 0.00 | 0.00 | 0.00 |
| Buyers | 0.48 | 0.48 | 0.03 | 0.00 | 0.00 | 0.00 | 0.00 | 0.00 | 0.00 | 0.00 | 0.00 | 0.00 |
| Investors | 0.34 | 0.32 | 0.34 | 0.00 | 0.00 | 0.00 | 0.00 | 0.00 | 0.00 | 0.00 | 0.00 | 0.00 |
| MatSteel | 0.00 | 0.00 | 0.00 | 1.00 | 0.00 | 0.00 | 0.00 | 0.00 | 0.00 | 0.00 | 0.00 | 0.00 |
| MatAl | 0.00 | 0.00 | 0.00 | 0.00 | 1.00 | 0.00 | 0.00 | 0.00 | 0.00 | 0.00 | 0.00 | 0.00 |
| UtlJP | 0.00 | 0.00 | 0.00 | 0.00 | 0.00 | 1.00 | 0.00 | 0.00 | 0.00 | 0.00 | 0.00 | 0.00 |
| UtlNA | 0.00 | 0.00 | 0.00 | 0.00 | 0.00 | 0.00 | 1.00 | 0.00 | 0.00 | 0.00 | 0.00 | 0.00 |
| UtlSA | 0.00 | 0.00 | 0.00 | 0.00 | 0.00 | 0.00 | 0.00 | 1.00 | 0.00 | 0.00 | 0.00 | 0.00 |
| UtlEU | 0.00 | 0.00 | 0.00 | 0.00 | 0.00 | 0.00 | 0.00 | 0.00 | 1.00 | 0.00 | 0.00 | 0.00 |
| UtlAO | 0.00 | 0.00 | 0.00 | 0.00 | 0.00 | 0.00 | 0.00 | 0.00 | 0.00 | 1.00 | 0.00 | 0.00 |
| UtlCH | 0.00 | 0.00 | 0.00 | 0.00 | 0.00 | 0.00 | 0.00 | 0.00 | 0.00 | 0.00 | 1.00 | 0.00 |
| RoW | 0.00 | 0.00 | 0.00 | 0.00 | 0.00 | 0.00 | 0.00 | 0.00 | 0.00 | 0.00 | 0.00 | 1.00 |

| | W | E | GHG |
|---|---|---|---|
| Unit | Mgal | TWh | MtCO2e |
| Honda | 7.63 | 3.57636 | 1.24 |
| Buyers | 0 | 972.95 | 234.4 |
| Investors | 0 | 0 | 0 |
| MatSteel | 2.7e+07 | 8290.5 | 10530.78 |
| MatAl | 366470 | 1930.61 | 1996.46 |
| UtlJP | 508670 | 997.4 | 417.51 |
| UtlNA | 2209520 | 4701.07 | 2983.31 |
| UtlSA | 280640 | 584.66 | 82.43 |
| UtlEU | 1490600 | 3635.6 | 1413.08 |
| UtlAO | 511630 | 1346.42 | 1030.95 |
| UtlCH | 1961900 | 3503.4 | 2959.85 |
| RoW | NA | NA | NA |

(c)

| | W | E | GHG |
|---|---|---|---|
| Unit | Mgal | TWh | MtCO2e |
| Honda | 7956.04 | 480.08 | 120.15 |
| Buyers | 7478.68 | 479.51 | 119.75 |
| Investors | 477.36 | 28.80 | 7.21 |
| MatSteel | 26989178.95 | 8288.13 | 10526.79 |
| MatAl | 365961.35 | 1928.24 | 1993.74 |
| UtlJP | 507059.55 | 995.31 | 416.45 |
| UtlNA | 2208197.97 | 4698.72 | 2981.63 |
| UtlSA | 280304.74 | 584.10 | 82.37 |
| UtlEU | 1490481.50 | 3635.38 | 1412.98 |
| UtlAO | 511225.75 | 1345.72 | 1030.22 |
| UtlCH | 1961115.74 | 3502.19 | 2958.71 |
| RoW | NA | NA | NA |

(d)

| | W | E | GHG |
|---|---|---|---|
| Unit | Mgal | TWh | MtCO2e |
| Honda | 3.8005 | 1.7814 | 0.6176 |
| Buyers | 0.0000 | 470.5584 | 113.3654 |
| Investors | 0.0000 | 0.0000 | 0.0000 |
| MatSteel | 5397.8348 | 1.6574 | 2.1053 |
| MatAl | 256.1691 | 1.3495 | 1.3956 |
| UtlJP | 811.2690 | 1.5907 | 0.6659 |
| UtlNA | 662.4572 | 1.4095 | 0.8945 |
| UtlSA | 168.1814 | 0.3504 | 0.0494 |
| UtlEU | 59.6192 | 0.1454 | 0.0565 |
| UtlAO | 204.4878 | 0.5381 | 0.4120 |
| UtlCH | 392.2225 | 0.7004 | 0.5917 |
| RoW | NA | NA | NA |
| Total | 7956.0416 | 480.0812 | 120.1540 |
| Change | 1042.7315 | 134.2374 | 96.8984 |

(e)

Table 7: The P-perspective MEIO model results for the Honda use case. a) The weights $w$, b) The responsibility $R$, c) The immediate footprint, d) The actual footprint, and e) The breakdown of the actual footprint of Honda based on its sources. Please note the change values are not in percentage.

| Source | Channel | List of econons | Value of econons | Non-zero Elements of w |
|---|---|---|---|---|
| Honda | 2 (Goods) | 1 vehicle (BuyerA),<br>22,185,000 vehicle (Buyers) | $20,000 (BuyerA),<br>$M107,479 (Buyers) | w(Honda, BuyerA) = 1.86 $10^{-5}$% |
| Honda | 4 (Money) | $M6,423 (Investors),<br>$M217.31 (MatSteel),<br>$M76.13 (MatAl),<br>$M247.04 (UtlJP),<br>$M107.49 (UtlNA),<br>$M32.23 (UtlSA),<br>$M15.84 (UtlEU),<br>$M84.99 (UtlAO),<br>$M43.50 (UtlCH),<br>$M100,231.5 (RoW) | $M6,423 (Investors),<br>$M217.31 (MatSteel),<br>$M76.13 (MatAl),<br>$M247.04 (UtlJP),<br>$M107.49 (UtlNA),<br>$M32.23 (UtlSA),<br>$M15.84 (UtlEU),<br>$M84.99 (UtlAO),<br>$M43.50 (UtlCH),<br>$M100,231.5 (RoW) | w(Honda, Investors) = 5.98%,<br>w(Honda, MatSteel) = 0.20%,<br>w(Honda, MatAl) = 0.07%,<br>w(Honda, UtlJP) = 0.23%,<br>w(Honda, UtlNA) = 0.1%,<br>w(Honda, UtlSA) = 0.03%,<br>w(Honda, UtlEU) = 0.01%,<br>w(Honda, UtlAO) = 0.08%,<br>w(Honda, UtlCH) = 0.04% |
| BuyerA | 4 (Money) | $20,000 (Honda),<br>3× $50,000 (RoW) | $20,000 (Honda),<br>3× $50,000 (RoW) | w(BuyerA, Honda) = 11.77% |
| Investors | All | — | — | w(Buyers, Honda) = 100% |
| MatSteel | 1 (Raw) | 241,459.50 t Steel (Honda),<br>1,499.76 Mt Steel (ROW) | $M217.31 (Honda),<br>$B1,349.78 (ROW) | w(MatSteel, Honda) = 0.02%,<br>w(MatSteel, ROW) = 99.98% |
| MatSteel | 4 (Money) | $M494.45 (UtlAO) | $M494.45 (UtlAO) | w(MatSteel, UtlAO) = 100% |
| MatAl | 1 (Raw) | 28,540.50 t Steel (Honda),<br>44.07 Mt Steel (ROW) | $M76.13 (Honda),<br>$M117,557.09 (ROW) | w(MatAl, Honda) = 0.07%,<br>w(MatAl, ROW) = 99.93% |
| MatAl | 4 (Money) | $M3,916.38 (UtlAO) | $M3,916.38 (UtlAO) | w(MatAl, UtlAO) = 100% |
| UtlJP | 1 (Raw) | 1.60 TWh Elect (Honda),<br>995.80 TWh Elect (ROW) | $M247.04 (Honda),<br>$M153,750 (ROW) | w(UtlJP, Honda) = 0.16%,<br>w(UtlJP, ROW) = 99.84% |
| UtlNA | 1 (Raw) | 1.54 TWh Elect (Honda),<br>4,699.53 TWh Elect (ROW) | $M107.49 (Honda),<br>$M328,030 (ROW) | w(UtlNA, Honda) = 0.03%,<br>w(UtlNA, ROW) = 99.97% |
| UtlSA | 1 (Raw) | 0.33 TWh Elect (Honda),<br>584.33 TWh Elect (ROW) | $M32.23 (Honda),<br>$M57,069.56 (ROW) | w(UtlSA, Honda) = 0.06%,<br>w(UtlSA, ROW) = 99.94% |
| UtlEU | 1 (Raw) | 0.15 TWh Elect (Honda),<br>3,635.45 TWh Elect (ROW) | $M15.84 (Honda),<br>$M383,903.5 (ROW) | w(UtlEU, Honda) = 0.004%,<br>w(UtlEU, ROW) = 99.996% |
| UtlAO | 1 (Raw) | 0.95 TWh Elect (Honda),<br>1,345.47 TWh Elect (ROW) | $M84.99 (Honda),<br>$M120,370 (ROW) | w(UtlAO, Honda) = 0.04%,<br>w(UtlAO, ROW) = 99.96% |
| UtlCH | 1 (Raw) | 0.58 TWh Elect (Honda),<br>3,502.82 TWh Elect (ROW) | $M43.50 (Honda),<br>$M262,711.5 (ROW) | w(UtlCH, Honda) = 0.02%,<br>w(UtlCH, ROW) = 99.98% |

Table 8: The gauge econons and also related responsibility graph's weights in the Honda E-perspective use case (footprint of a typical buyer).

contribution.

Secondly, if we compare the BuyerA's footprint with the dividend share footprint of a typical buyer using the footprint estimated for the Buyers entity using the P-perspective MEIO model in the previous section, the E-perspective footprint of BuyerA would be much bigger. We justify that the second case, i.e., E-perspective footprint, is more credible because the buyers are the main motive force that make the service/product providers proceed with their business and actions. The *sector slicing hyper-line* introduced in the framework provides the required mechanism that prevents over estimation of the end uses' responsibility when we calculate the footprint of the provider. In contrast, when we estimate the footprint of a single buyer, a higher footprint actually helps to rule out any concern about footprint under-estimation. In addition, the E-perspective MEIO modeling provides a mechanism to estimate the footprint of buyers in lower granularities, and this is one of the advantages of the proposed modified MEIO frameworks compared to its naive version. To put this into numbers, the estimated GHG footprint of BuyerA in the E-perspective, which is 152.60 Kg$CO_2$e, is 16.40 times bigger than the average GHG footprint per buyer calculated from the P-perspective results, which would be 9.30 Kg$CO_2$e.[9] In other words, the E-perspective MEIO modeling ensures that an individual's footprint is not undermined because of *sharing* phenomenon when a big ensemble of individuals participate in the same action. In contrast, the P-perspective works toward blocking leakage of the footprint from the providers to the

[9] Considering half footprint for motorcycles compared to cars, and using the values of Tables S.7 and 7(d). The rest of 18.21 Kg$CO_2$e immediate footprint of a single car, i.e., 8.91 Kg$CO_2$e, was transferred to Honda in the P-perspective model.

(a)

| | Honda | BuyerA | Investors | MatSteel | MatAl | UtlJP | UtlNA | UtlSA | UtlEU | UtlAO | UtlCH | RoW |
|---|---|---|---|---|---|---|---|---|---|---|---|---|
| Honda | 100.00 | 0.00 | 6.00 | 0.20 | 0.07 | 0.23 | 0.10 | 0.03 | 0.02 | 0.08 | 0.04 | 0.00 |
| BuyerA | 11.77 | 100.00 | 0.00 | 0.00 | 0.00 | 0.00 | 0.00 | 0.00 | 0.00 | 0.00 | 0.00 | 0.00 |
| Investors | 100.00 | 0.00 | 100.00 | 0.00 | 0.00 | 0.00 | 0.00 | 0.00 | 0.00 | 0.00 | 0.00 | 0.00 |
| MatSteel | 0.02 | 0.00 | 0.00 | 100.00 | 0.00 | 0.00 | 0.00 | 0.00 | 0.00 | 0.00 | 0.00 | 0.00 |
| MatAl | 0.07 | 0.00 | 0.00 | 0.00 | 100.00 | 0.00 | 0.00 | 0.00 | 0.00 | 0.00 | 0.00 | 0.00 |
| UtlJP | 0.16 | 0.00 | 0.00 | 0.00 | 0.00 | 100.00 | 0.00 | 0.00 | 0.00 | 0.00 | 0.00 | 0.00 |
| UtlNA | 0.03 | 0.00 | 0.00 | 0.00 | 0.00 | 0.00 | 100.00 | 0.00 | 0.00 | 0.00 | 0.00 | 0.00 |
| UtlSA | 0.06 | 0.00 | 0.00 | 0.00 | 0.00 | 0.00 | 0.00 | 100.00 | 0.00 | 0.00 | 0.00 | 0.00 |
| UtlEU | 0.00 | 0.00 | 0.00 | 0.00 | 0.00 | 0.00 | 0.00 | 0.00 | 100.00 | 0.00 | 0.00 | 0.00 |
| UtlAO | 0.04 | 0.00 | 0.00 | 0.00 | 0.00 | 0.00 | 0.00 | 0.00 | 0.00 | 100.00 | 0.00 | 0.00 |
| UtlCH | 0.02 | 0.00 | 0.00 | 0.00 | 0.00 | 0.00 | 0.00 | 0.00 | 0.00 | 0.00 | 100.00 | 0.00 |
| RoW | 0.00 | 0.00 | 0.00 | 0.00 | 0.00 | 0.00 | 0.00 | 0.00 | 0.00 | 0.00 | 0.00 | 0.00 |

(b)

| | Honda | BuyerA | Investors | MatSteel | MatAl | UtlJP | UtlNA | UtlSA | UtlEU | UtlAO | UtlCH | RoW |
|---|---|---|---|---|---|---|---|---|---|---|---|---|
| Honda | 93.66 | 0.00 | 5.62 | 0.19 | 0.07 | 0.22 | 0.09 | 0.03 | 0.01 | 0.07 | 0.04 | 0.00 |
| BuyerA | 10.46 | 88.84 | 0.63 | 0.02 | 0.01 | 0.02 | 0.01 | 0.00 | 0.00 | 0.01 | 0.00 | 0.00 |
| Investors | 49.81 | 0.00 | 49.81 | 0.10 | 0.04 | 0.11 | 0.05 | 0.01 | 0.01 | 0.04 | 0.02 | 0.00 |
| MatSteel | 0.02 | 0.00 | 0.00 | 99.98 | 0.00 | 0.00 | 0.00 | 0.00 | 0.00 | 0.00 | 0.00 | 0.00 |
| MatAl | 0.07 | 0.00 | 0.00 | 0.00 | 99.93 | 0.00 | 0.00 | 0.00 | 0.00 | 0.00 | 0.00 | 0.00 |
| UtlJP | 0.16 | 0.00 | 0.01 | 0.00 | 0.00 | 99.83 | 0.00 | 0.00 | 0.00 | 0.00 | 0.00 | 0.00 |
| UtlNA | 0.03 | 0.00 | 0.00 | 0.00 | 0.00 | 0.00 | 99.97 | 0.00 | 0.00 | 0.00 | 0.00 | 0.00 |
| UtlSA | 0.06 | 0.00 | 0.00 | 0.00 | 0.00 | 0.00 | 0.00 | 99.94 | 0.00 | 0.00 | 0.00 | 0.00 |
| UtlEU | 0.00 | 0.00 | 0.00 | 0.00 | 0.00 | 0.00 | 0.00 | 0.00 | 100.00 | 0.00 | 0.00 | 0.00 |
| UtlAO | 0.04 | 0.00 | 0.00 | 0.00 | 0.00 | 0.00 | 0.00 | 0.00 | 0.00 | 99.96 | 0.00 | 0.00 |
| UtlCH | 0.02 | 0.00 | 0.00 | 0.00 | 0.00 | 0.00 | 0.00 | 0.00 | 0.00 | 0.00 | 99.98 | 0.00 |
| RoW | 0.00 | 0.00 | 0.00 | 0.00 | 0.00 | 0.00 | 0.00 | 0.00 | 0.00 | 0.00 | 0.00 | 100.00 |

| | W | E | GHG |
|---|---|---|---|
| Unit | kgal | GWh | ktCO2e |
| Honda | 7630 | 3576.36 | 1240 |
| BuyerA | 0 | 75.59 | 0.01821 |
| Investors | 0 | 0 | 0 |
| MatSteel | 2.7e+10 | 8290500 | 10530780 |
| MatAl | 366470000 | 1930610 | 1996460 |
| UtlJP | 508670000 | 997400 | 417510 |
| UtlNA | 2209520000 | 4701070 | 2983310 |
| UtlSA | 280640000 | 584660 | 82430 |
| UtlEU | 1490600000 | 3635600 | 1413080 |
| UtlAO | 511630000 | 1346420 | 1030950 |
| UtlCH | 1961900000 | 3503400 | 2959850 |
| RoW | NA | NA | NA |

(c)

| | W | E | GHG |
|---|---|---|---|
| Unit | kgal | GWh | ktCO2e |
| Honda | 7962222.93 | 11103.61 | 7335.18 |
| BuyerA | 148.10 | 67.36 | 0.15 |
| Investors | 477733.38 | 666.22 | 440.11 |
| MatSteel | 26994251742.55 | 8288752.28 | 10528546.47 |
| MatAl | 366201969.63 | 1929176.08 | 1994974.23 |
| UtlJP | 507820846.21 | 995724.61 | 416814.86 |
| UtlNA | 2208820478.08 | 4699575.83 | 2982362.08 |
| UtlSA | 280462725.42 | 584289.04 | 82379.43 |
| UtlEU | 1490537538.38 | 3635446.41 | 1413020.75 |
| UtlAO | 511417882.85 | 1345854.00 | 1030515.70 |
| UtlCH | 1961484342.49 | 3502656.51 | 2959221.04 |
| RoW | NA | NA | NA |

(d)

| | W | E | GHG |
|---|---|---|---|
| Unit | kgal | GWh | ktCO2e |
| Honda | 0.1329 | 0.0623 | 0.0216 |
| BuyerA | 0.0000 | 67.1516 | 0.0162 |
| Investors | 0.0000 | 0.0000 | 0.0000 |
| MatSteel | 100.4186 | 0.0308 | 0.0392 |
| MatAl | 4.7679 | 0.0251 | 0.0260 |
| UtlJP | 15.1123 | 0.0296 | 0.0124 |
| UtlNA | 12.3252 | 0.0262 | 0.0166 |
| UtlSA | 3.1299 | 0.0065 | 0.0009 |
| UtlEU | 1.1090 | 0.0027 | 0.0011 |
| UtlAO | 3.8049 | 0.0100 | 0.0077 |
| UtlCH | 7.2967 | 0.0130 | 0.0110 |
| RoW | NA | NA | NA |
| Total | 148.0973 | 67.3580 | 0.1526 |
| Change | Inf | 0.8911 | 8.3806 |

(e)

Table 9: The E-perspective MEIO model results for the Honda use case. a) The weights $w$, b) The responsibility $R$, c) The immediate footprint, d) The actual footprint, and e) The breakdown of the actual footprint of a typical buyer. Please note the change values are not in percentage.

| | UK Grid | Total Footprint |
|---|---|---|
| GHG Factor (KgCO$_2$e/MWh) | 476.83 | |
| Water Factor (Gal/kWh) | 0.53 | |
| Conversion Factor (MWh/GJ) | 0.28 | |
| BT UK Energy Consumption (GJ) | 8,424,000.00 | |
| BT UK Energy Footprint (GWh) | 2,340.19 | 2,340.19 |
| BT UK Energy-related GHG Footprint (Scope 1) (KtCO$_2$e) | 1,124.71 | ~~1,124.71~~ |
| BT UK GHG Footprint (Scope 1) (KtCO$_2$e) | | 169.18 |
| BT UK GHG Footprint Total (KtCO$_2$e) | | 169.18 |
| BT UK Energy-related Water Footprint (Scope 1) (KGal) | 1,240.30 | ~~1,240.30~~ |
| BT UK Water Footprint Tap (MGal) | | 364.00 |
| BT UK Water Footprint Ground (MGal) | | 0.00 |
| BT UK Water Footprint Total (MGal) | | 364.00 |

Table 10: The *immediate* BT UK's footprint. The water consumption for on-site electricity generation is not summed because we were not sure it has been already accounted for. The Liter unit in the footprint factors refer to the fuel used.

end users, while allowing the footprint of end users be shared by the providers in a responsibility-based approach, instead of a direct summation.

### *4.1.3. Post Use Case Discussions*

It is worth mentioning that there are other entities that could be added into the ecosystem picture of this use case. For example, the gasoline distributors and their associated fuel-transportation footprint could be considered. In such a case, the end users will interact with two providers, i.e., the Auto maker and the fuel distributors, and the slicing hyper-line should be accordingly updated. Furthermore, in another case, the whole network of fuel distributors can be considered as 'Telco'-like ecosystem if the amount and time of end users fueling is recorded and is available. This could be leveraged toward dynamic behavioral promotions to reduce the transport footprint, fuel leakage at interfaces, for example, the interface of the gasoline pump and the car's tank. and potentially reduce the actual fuel consumption of the end users.

Another aspect that could be potentially included in the modeling is the role of the provider's employee. The employees could be added to the models in a way similar to the end users. In this case, another slicing hyper-line would allow the Channel 4's, i.e., money, interactions transferred between the provider and the employee.

## *4.2. BT Use case*

The union set of the EoI also the three end user entities of this use case is: **BTUK** The BT UK, **ClientsBB** The broadband-service clients, **ClientsPh** The phone-service clients, **ClientsTV** The TV-service clients, **Investors** The BT's investors, **Community** The community received contribution from BT, **UtlUK** The electricity utilities in UK, **Ads** The advertisement agencies. As can be seen, there are three entities that represent the end users. Also, a period of one month is considered as $\Delta t$. The immediate footprint of BT UK is summarized in Table 10.

Also, the P-perspective econons and the final responsibility and footprint of the BT UK use case are provided in Tables 11 and 12. The breakdown of the BT UK footprint , Table 12(e), shows that their footprint is moderately higher than their immediate footprint with a more tendency toward electricity-related footprint.

| Source | Channel | List of econons | Value of econons | Non-zero Elements of $w$ |
|---|---|---|---|---|
| BTUK | 3 (Service) | 9,174.22 TB/m (ClientsBB),<br>4,152.83 Mminutes/m (ClientsPh),<br>69.69 Mhours/m (ClientsTV) | &M28.25 (ClientsBB),<br>&M392.83 (ClientsPh),<br>&M28.93 (ClientsTV) | $w$(BTUK, ClientsBB) = 1.69%,<br>$w$(BTUK, ClientsPh) = 23.48%,<br>$w$(BTUK, ClientsTV) = 1.73% |
| BTUK | 4 (Money) | &M45.25 (Investors),<br>&M2.66 (Community),<br>&M20.83 (UtlUK) | &M45.25 (Investors),<br>&M2.66 (Community),<br>&M20.83 (UtlUK) | $w$(BTUK, Investors) = 2.71%,<br>$w$(BTUK, Community) = 0.16%,<br>$w$(BTUK, UtlUK) = 1.25% |
| ClientsBB | 4 (Money) | &M28.25 (BTUK) | &M28.25 (BTUK) | $w$(ClientsBB, BTUK) = 2.36% |
| ClientsPh | 4 (Money) | &M392.83 (BTUK) | &M392.83 (BTUK) | $w$(ClientsBB, BTUK) = 1.05% |
| ClientsTV | 4 (Money) | &M28.93 (BTUK) | &M28.93 (BTUK) | $w$(ClientsBB, BTUK) = 2.29% |
| Investors | All | — | — | $w$(Investors, BTUK) = 100% |
| UtlUK | 1 (Raw) | 195 GWh Elect (BTUK),<br>29.12 TWh Elect (ROW) | | $w$(UtlJP, ROW) = 99.33% |
| UtlUK | 4 (Money) | &M20.83 (BTUK),<br>&B2.20 (ROW) | &M20.83 (BTUK),<br>&B2.20 (ROW) | $w$(UtlUK, BTUK) = 0.95% |
| Ads | 4 (Money) | &M12.07 (BTUK),<br>&M290 (ROW) | &M12.07 (BTUK),<br>&M290 (ROW) | $w$(Ads, BTUK) = 4%,<br>$w$(Ads, ROW) = 96% |

Table 11: The aggregated monthly gauge econons and also related responsibility graph's weights in the BT UK P-perspective use case.

In addition, it is worth mentioning that the responsibility of the Ads entity could not be calculated with the current graph. This will be addressed in the future by adding other entities and econons that would represent the transfers to the Ads entity.

If the responsibility and footprint of a particular client is required, the E-perspective MEIO model can be used. We consider a typical ClientsBB subscriber and denoted them as CliBBA. The associated gauge econons, weights, responsibilities, and footprint are presented in Tables 13 and 14.

### *4.3. A Note on the relation between of MEIO models and LCA methods*

The use cases provided here have shown that the MEIO approach can simply handle the Scope-3 footprint associated to an actor by considering their responsibility for those actions incur outside their direct control.[10] In addition, it is easy to observe that this inclusion does not require that the actions should be simultaneous. Therefore 'upstream' and 'downstream' footprint, which are of the interest in the LCA approaches, can be enveloped by considering proper time intervals. In other words, the MEIO approach can be considered as a converging means that could allow integration of both Scope-3 and LCA perspectives in the same model.

It is also worth adding that many other approaches have been considered to integrate Input-Output (IO) methods with the LCA approaches. For example, in Huang et al. (2011, 2009), an Economic Input-Output-LCA (EIO-LCA) model was introduced using a matrix model. In another work on calculating the Scope-3 footprint in Australia (Downie and Stubbs, 2012, 2013), it was concluded that the LCA approaches are not able to capture the Scope 3, and it was suggested to consider them along with the proper IO methods. Other examples are (Busch, 2011; Meinrenken et al., 2012; Murray et al., 2011), which also promote utilization of the IO approaches in the form of hybrid EIO-LCA methodologies (Ewing et al., 2011).

Another concern about the LCA analyses, which are very important in assessing the level of nature-friendliness of a provider, raises when we want to calculate the share of the end users. We argue that considering only the operational footprint and the end users share in that footprint is more fair toward

[10] An example of the Scope-3 footnote could be the GHG emissions associated to the employees' commute (Shrake et al., 2011).

(a)

| | BTUK | ClientsBB | ClientsPh | ClientsTV | Investors | Community | UtlUK | Ads | RoW |
|---|---|---|---|---|---|---|---|---|---|
| BTUK | 1.00 | 0.02 | 0.23 | 0.02 | 0.03 | 0.00 | 0.01 | 0.00 | 0.00 |
| ClientsBB | 1.00 | 1.00 | 0.00 | 0.00 | 0.00 | 0.00 | 0.00 | 0.00 | 0.00 |
| ClientsPh | 1.00 | 0.00 | 1.00 | 0.00 | 0.00 | 0.00 | 0.00 | 0.00 | 0.00 |
| ClientsTV | 1.00 | 0.00 | 0.00 | 1.00 | 0.00 | 0.00 | 0.00 | 0.00 | 0.00 |
| Investors | 1.00 | 0.00 | 0.00 | 0.00 | 1.00 | 0.00 | 0.00 | 0.00 | 0.00 |
| Community | 1.00 | 0.00 | 0.00 | 0.00 | 0.00 | 1.00 | 0.00 | 0.00 | 0.00 |
| UtlUK | 0.01 | 0.00 | 0.00 | 0.00 | 0.00 | 0.00 | 1.00 | 0.00 | 0.00 |
| Ads | 0.04 | 0.00 | 0.00 | 0.00 | 0.00 | 0.00 | 0.00 | 1.00 | 0.00 |
| RoW | 0.00 | 0.00 | 0.00 | 0.00 | 0.00 | 0.00 | 0.00 | 0.00 | 0.00 |

(b)

| | BTUK | ClientsBB | ClientsPh | ClientsTV | Investors | Community | UtlUK | Ads | RoW |
|---|---|---|---|---|---|---|---|---|---|
| BTUK | 0.76 | 0.01 | 0.18 | 0.01 | 0.02 | 0.00 | 0.01 | 0.00 | 0.00 |
| ClientsBB | 0.44 | 0.44 | 0.10 | 0.01 | 0.01 | 0.00 | 0.01 | 0.00 | 0.00 |
| ClientsPh | 0.48 | 0.01 | 0.48 | 0.01 | 0.01 | 0.00 | 0.01 | 0.00 | 0.00 |
| ClientsTV | 0.44 | 0.01 | 0.10 | 0.44 | 0.01 | 0.00 | 0.01 | 0.00 | 0.00 |
| Investors | 0.44 | 0.01 | 0.10 | 0.01 | 0.44 | 0.00 | 0.01 | 0.00 | 0.00 |
| Community | 0.43 | 0.01 | 0.10 | 0.01 | 0.01 | 0.43 | 0.01 | 0.00 | 0.00 |
| UtlUK | 0.01 | 0.00 | 0.00 | 0.00 | 0.00 | 0.00 | 0.99 | 0.00 | 0.00 |
| Ads | 0.04 | 0.00 | 0.01 | 0.00 | 0.00 | 0.00 | 0.00 | 0.95 | 0.00 |
| RoW | 0.00 | 0.00 | 0.00 | 0.00 | 0.00 | 0.00 | 0.00 | 0.00 | 1.00 |

| | W | E | GHG |
|---|---|---|---|
| Unit | Mgal | GWh | ktCO2e |
| BTUK | 30.333 | 195.01583 | 52.83 |
| ClientsBB | 61.39 | 115.83 | 137.51 |
| ClientsPh | 2.20 | 4.16 | 4.94 |
| ClientsTV | 3.36 | 6.34 | 7.85 |
| Investors | 0.00 | 0.00 | 0.00 |
| Community | 0.00 | 0.00 | 0.00 |
| UtlUK | 15537.5 | 29316.667 | 34792.5 |
| Ads | 0.00 | 0.00 | 0.00 |
| RoW | NA | NA | NA |

(c)

| | W | E | GHG |
|---|---|---|---|
| Unit | Mgal | GWh | ktCO2e |
| BTUK | 198.25 | 479.24 | 432.59 |
| ClientsBB | 29.67 | 57.75 | 66.26 |
| ClientsPh | 47.36 | 114.06 | 103.39 |
| ClientsTV | 4.87 | 11.01 | 10.85 |
| Investors | 5.37 | 12.99 | 11.72 |
| Community | 0.32 | 0.77 | 0.69 |
| UtlUK | 15348.94 | 28962.20 | 34370.12 |
| Ads | NA | NA | NA |
| RoW | NA | NA | NA |

(d)

| | W | E | GHG |
|---|---|---|---|
| Unit | Mgal | GWh | ktCO2e |
| BTUK | 23.15 | 148.84 | 40.32 |
| ClientsBB | 26.77 | 50.51 | 59.96 |
| ClientsPh | 1.06 | 2.00 | 2.38 |
| ClientsTV | 1.47 | 2.77 | 3.42 |
| Investors | 0.00 | 0.00 | 0.00 |
| Community | 0.00 | 0.00 | 0.00 |
| UtlUK | 145.81 | 275.12 | 326.50 |
| Ads | 0.00 | 0.00 | 0.00 |
| RoW | NA | NA | NA |
| Total | 198.25 | 479.24 | 432.59 |
| Change | 6.54 | 2.46 | 8.19 |

(e)

Table 12: The P-perspective MEIO model results for the BT UK use case. a) The weights $w$, b) The responsibility $R$, c) The immediate footprint, d) The actual footprint, and e) The breakdown of the actual footprint of BT UK based on its sources. Please note the change values are not in percentage.

| Source | Channel | List of econons | Value of econons | Non-zero Elements of $w$ |
|---|---|---|---|---|
| BTUK | 3 (Service) | 17 TB/m (CliBBA) | &52.36 (CliBBA) | $w$(BTUK, CliBBA) = 2.61 $10^{-7}$% |
| BTUK | 4 (Money) | &M45.25 (Investors), &M2.66 (Community), &M20.83 (UtlUK) | &M45.25 (Investors), &M2.66 (Community), &M20.83 (UtlUK) | $w$(BTUK, Investors) = 2.71%, $w$(BTUK, Community) = 0.16%, $w$(BTUK, UtlUK) = 1.25% |
| CliBBA | 4 (Money) | &52.36 (BTUK) | &M52.36 (BTUK) | $w$(ClientsBB, BTUK) = 2.37% |
| Investors | All | — | — | $w$(Investors, BTUK) = 100% |
| UtlUK | 1 (Raw) | 195 GWh Elect (BTUK), 29.12 TWh Elect (ROW) | | $w$(UtlJP, ROW) = 99.33% |
| UtlUK | 4 (Money) | &M20.83 (BTUK), &B2.20 (ROW) | &M20.83 (BTUK), &B2.20 (ROW) | $w$(UtlUK, BTUK) = 0.95% |
| Ads | 4 (Money) | &M12.07 (BTUK), &M290 (ROW) | &M12.07 (BTUK), &M290 (ROW) | $w$(Ads, BTUK) = 4%, $w$(Ads, ROW) = 96% |

Table 13: The gauge econons and also related responsibility graph's weights in the BT UK E-perspective use case (footprint of a typical ClientBB subscriber).

(a)

| | BTUK | CliBBA | Investors | Community | UtlUK | Ads | RoW |
|---|---|---|---|---|---|---|---|
| BTUK | 100.00 | 0.00 | 2.71 | 0.16 | 1.25 | 0.00 | 0.00 |
| CliBBA | 2.37 | 100.00 | 0.00 | 0.00 | 0.00 | 0.00 | 0.00 |
| Investors | 100.00 | 0.00 | 0.00 | 0.00 | 0.00 | 0.00 | 0.00 |
| Community | 100.00 | 0.00 | 0.00 | 0.00 | 0.00 | 0.00 | 0.00 |
| UtlUK | 0.95 | 0.00 | 0.00 | 0.00 | 0.00 | 0.00 | 0.00 |
| Ads | 4.00 | 0.00 | 0.00 | 0.00 | 0.00 | 0.00 | 0.00 |
| RoW | 0.00 | 0.00 | 0.00 | 0.00 | 0.00 | 0.00 | 0.00 |

(b)

| | BTUK | CliBBA | Investors | Community | UtlUK | Ads | RoW |
|---|---|---|---|---|---|---|---|
| BTUK | 96.04 | 0.00 | 2.60 | 0.15 | 1.20 | 0.00 | 0.00 |
| CliBBA | 2.31 | 97.59 | 0.06 | 0.00 | 0.03 | 0.00 | 0.00 |
| Investors | 49.65 | 0.00 | 49.65 | 0.08 | 0.62 | 0.00 | 0.00 |
| Community | 49.03 | 0.00 | 1.33 | 49.03 | 0.61 | 0.00 | 0.00 |
| UtlUK | 0.94 | 0.00 | 0.03 | 0.00 | 99.03 | 0.00 | 0.00 |
| Ads | 3.84 | 0.00 | 0.10 | 0.01 | 0.05 | 96.00 | 0.00 |
| RoW | 0.00 | 0.00 | 0.00 | 0.00 | 0.00 | 0.00 | 100.00 |

| | W | E | GHG |
|---|---|---|---|
| Unit | gal | MWh | tCO2e |
| BTUK | 30333000 | 195015830 | 52830000 |
| CliBBA | 12.91462 | 24.36591 | 25.49002 |
| Investors | 0 | 0 | 0 |
| Community | 0 | 0 | 0 |
| UtlUK | 15537500000 | 29316667000 | 34792500000 |
| Ads | 0 | 0 | 0 |
| RoW | NA | NA | NA |

(c)

| | W | E | GHG |
|---|---|---|---|
| Unit | gal | MWh | tCO2e |
| BTUK | 175310430.54 | 463111973.17 | 378069388.13 |
| CliBBA | 58.36 | 144.65 | 123.55 |
| Investors | 4750912.67 | 12550334.47 | 10245680.42 |
| Community | 280496.69 | 740979.16 | 604911.02 |
| UtlUK | 15387491114.66 | 29035279422.92 | 34456409922.37 |
| Ads | NA | NA | NA |
| RoW | NA | NA | NA |

(d)

| | W | E | GHG |
|---|---|---|---|
| Unit | gal | MWh | tCO2e |
| BTUK | 7.60 | 48.89 | 13.24 |
| CliBBA | 12.60 | 23.78 | 24.88 |
| Investors | 0.00 | 0.00 | 0.00 |
| Community | 0.00 | 0.00 | 0.00 |
| UtlUK | 38.15 | 71.99 | 85.43 |
| Ads | 0.00 | 0.00 | 0.00 |
| RoW | NA | NA | NA |
| Total | 58.36 | 144.65 | 123.55 |
| Change | 4.52 | 5.94 | 4.85 |

(e)

Table 14: The E-perspective MEIO model results for the BT UK use case. a) The weights $w$, b) The responsibility $R$, c) The immediate footprint, d) The actual footprint, and e) The breakdown of the actual footprint of a typical ClientBB subscriber. Please note the change values are not in percentage

the end users. For example, if a provider is performing recycling on its components at the moment, we cannot praise the current end users because the recycled components are not those which are used to serve the current end users. At the same time, the end uses cannot be charged with end-of-life footprint if their service provider does not have a recycling process in place at the moment, because it is possible that the provider establishes such processes in the future when the components used to serve the current users reach their end of life. In contrast, the MEIO approach brings the end users in the 'game' and let them decide if they are fine with footprint allocated to them based on their responsibility in the business drive of the provider, or they prefer to switch to another provider with less footprint allocated to the end users.

Also, in terms of sustainability and green projects which a provider may consider in order to reduce their footprint, it is suggested to plan for such projects within highly-low granular regions or 'communities.' This is especially important for small-scale green project, which otherwise would not be noticeable in the final footprint of the end users. When realized within a specific community, the green project impact on *that* community's footprint could be easily estimated and highlighted to the community and the rest of the end users. This can be performed by considering a specific additional entity in the MEIO model to represent the end users who belong to that community. This not only enables *rewarding* the participating end users, it also promotes similar projects in other regions or communities. We will consider estimating of the impact of such projects, such as the GreenStar Project (Farrahi Moghaddam et al., 2012a), on the footprint of small communities using the MEIO models in another work.

## 5. Conclusion and future prospects

A new Input-Output model, called the Multi-Entity Input-Output (MEIO) model, has been proposed to estimate the responsibility and footprint of entities. The proposed MEIO modeling approach, which is along the Everybody-in-the-Loop (EitL) framework, allows deploying mechanisms to promote behavioral changes, such as green or sustainability footprint bill mechanism. First, a naive MEIO model, has been introduced, which is indifference with respect to the end users. The core of the proposed MEIO model is a graph-based representation of entities and their actions that is used to calculate the unnormalized responsibility and also the final responsibility of every actor, and then to re-allocate immediate footprint of actors to those who are implicitly responsible. In terms of the footprint categories, we considered three footprint categories of Water, Energy, and GHG emissions. Adding other categories, such as black carbon, will be considered in future to make the models aware of natural and social impacts, such as air pollution, which could be contained to short time intervals and also to contained geographical regions. The naive model has been then generalized to P-perspective and E-perspective MEIO models in order to make it applicable to cases where a large number of end users are served by a provider. In particular, the E-perspective models are used to estimate the personalized footprint associated to a specific end user. In two use cases from the Auto and Telco industries, the advantages and potential of the proposed MEIO models have been illustrated, particularly on how to avoid footprint leakage to the end users. Another advantage of the MEIO models is their capability to implicitly provide some features of the Scope-3 and also LCA approaches, and potentially server as an integration candidate for such concepts.

As mentioned in the text, the data used in the use cases could be incomplete, and it was only used for the purpose of illustrating the MEIO models in practice. Re-estimating the footprint shares and responsibilities will be considered in future work using more comprehensive validated data from the providers and other sources. Also, generalization of the models in a hierarchical methodology will be considered to enable the MEIO models to work across various time scales and intervals. These hierarchical models would also help in smoothing the impact of fluctuation that could happen in small time intervals. When considering very big time intervals, the concept of *capacity* will be considered to add basic memory capabilities to the models.

Finally, developing behavioral analysis models to study and implement behavioral changes using the footprint bills generated using the MEIO models is an important future prospect that would require a multi-objective approach to optimization. This is along the prospected generalizations to the MEIO models in order to make them capable to forecast and predict the future interactions and gauge econons transferred among the entities. Various behavioral modeling strategies will be considered including the Fuzzy Cognitive Maps (FCMs).

### Acknowledgments

The authors thank the NSERC of Canada for their financial support under Grant CRDPJ 424371-11 and also under the Canada Research Chair in Sustainable Smart Eco-Cloud.

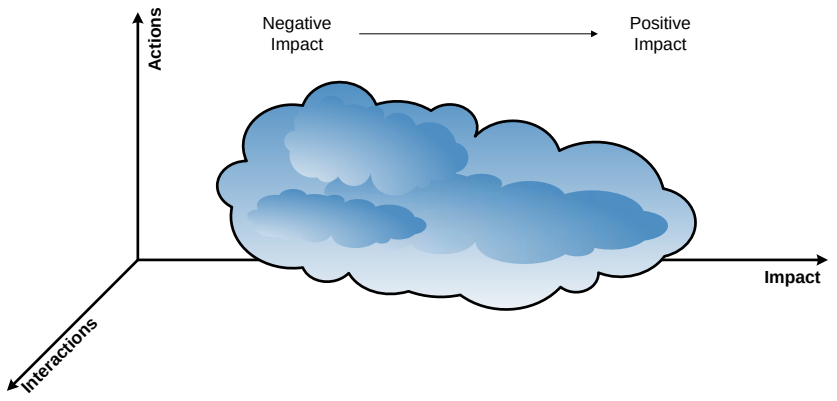

(a) The reference behavior.

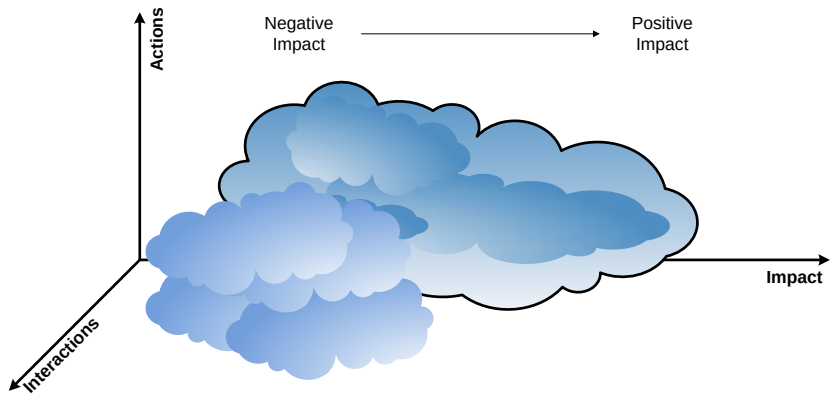

(b) An Incabehavior with negative impact.

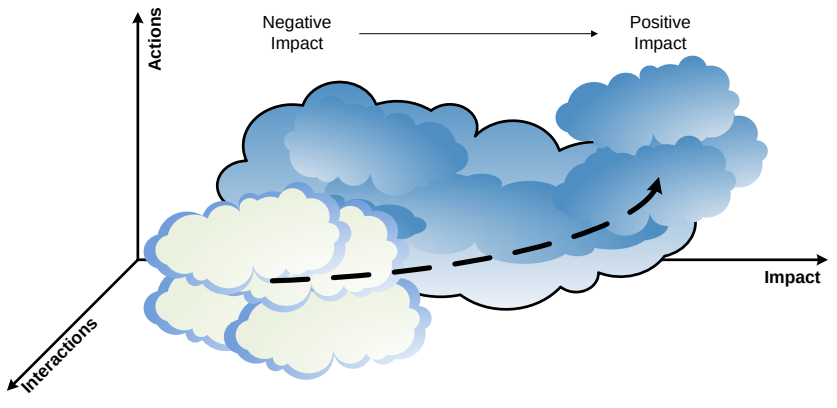

(c) Excabehavior with positive impact, compared to (b).

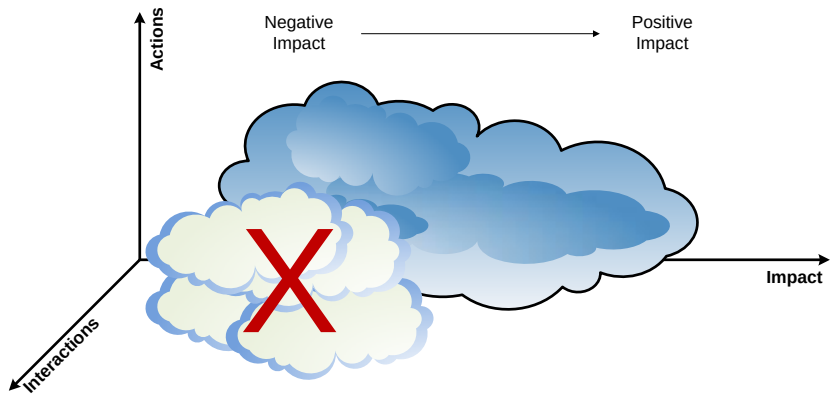

(d) Decabehavior with positive impact, compared to (b).

Figure A.1: The behavior cloud.

## Appendix A. Actors' Behavior and Sustainability

### *Appendix A.1. Behavior Modeling and Behavior Modification*

It has been observed by many that motivating an entity to change its behavior is much more effective than imposing regulations on them (Brynjarsdóttir et al., 2012; Lewis, 1993; Markard et al., 2012). From the sustainability perspective, this requires a management mechanism that targets induction of 'changes' in the end users behavior to make their lives, and therefore the world, more sustainable. The most-known behavior changes are pro-actions, such as recycling (Tangsubkul et al., 2005). This type of actions are very rewarding because the actor feels being distinctive and active. In contrast, concepts such as gray behavior (Ferebee, 2010) and negabehavior (Ross and Tomlinson, 2011) encourage the actors to pass on and reduce their actions, and adapt to 'the flow.' A combination of pro- and nega-behavior can be seen in the form of substituting habitual actions with more sustainable actions.

We reformulate these concepts in a form of a behavior model as shown in Figure A.1. The behavior space is modeled along three dimension of actions, interactions, and impact. A particular entity's behavior can be imagined as a cloud within this space that represent the likelihood of their behavior. In this model, a behavior change stands in one of three categories: i) deca-behavior, ii) inca-behavior, and iii) exca-behavior, which represent increase, decrease, or exchange of behavior, respectively. At the same time, the outcome of a particular behavior change depends on its start and end behaviors. For example, in Figure A.1(c), a excabehavior with a 'positive' impact is shown. In other words, in contrast to the general believe, decabehavior and excabehavior do not always implicitly carry a positive impact.

In terms of the practical tools toward sustainability, the ICT has been identified as an enabler toward promoting sustainable changes (Heeks, 2008; Petrov, 2008; Romm and Taylor, 2000). However, the ICT ecosystem itself is also required to participate in such changes. The Everybody in the Loop (EitL) framework and similar approaches, which promote providing sustainability statement to all actors and entities, are promising approaches to positively engaging everybody even in the case of highly heterogeneous ecosystems, such as that of the ICT. It is suggested that the sustainability statements provide various preselected suggestions in the form of deca-, inca-, and exca-behavior changes in addition to the estimated footprint in order to promote collective behavior changes.

*Appendix A.2. The Everybody in the Loop (EitL) Approach*

The EitL framework is not the first attempt to put the users in the loop in the form of a negative feedback. For example, in Schoenen et al. (2012, 2011), the user-in-the-loop (UitL) approach was introduced. Although the UitL approach is very interesting, it implicitly assumes that the end users and service providers form a 'system.' This systemic picture is valid when the service providers *govern* the interactions, i.e., when there is a non-competitive market and also the providers are not *selfish*. However, in the real world and in competitive markets, the service providers also act as selfish actors toward increasing their profit. This means that excluding the providers from the model may cause the end users would be overcharged or their shopping right would be compromised, and therefore an ecosystemic picture is more suitable (Farrahi Moghaddam et al., 2012b). The EitL approach is one way to consider all actors regardless of their size. The proposed EitL approach tries to include all actors of the target ecosystem, which its borders are placed such that there is little interactions and transfer across them. The EitL's subsequential transparency can be actually used by the service providers in order to diagnosis weak-links within themselves and also with other entities.

## Appendix B. Gauge Econons: Interaction Objects in the MEIO Models

Although most of financial interactions could be modeled in the form of transactions (Argyres and Zenger, 2012; Gedajlovic and Carney, 2010; Persson, 2010; Wagner, 1995),[11] with the increase in the complexity of the systems and interactions in recent years, many interactions and transfers do not fit within a standard transactional picture. In particular, many interactions would be actually 'unidirectional' transfers with either 'missing' or misplaced opposite-direction balancing transfers in unbounded time intervals, physical distances, or non-physical layers. This is our motivation to introduce gauge econons here to represent each interaction and transfer itself without imposing any transactional requirement.

In particular, all interactions between or among entities in the MEIO models are modeled by a set of 'collocated' interactions. Here, collocation is assumed to be beyond its physical location limitations. Even transactions can be imagined as a form of interaction comprised of a set of two collocated actions between two entities. First, in a new definition of transaction, we relax the requirement of being an exchange of an asset 'for payment.' Therefore, the proposed relaxed definition is closer to the original practice of transaction in the past centuries ago, and it would be general enough to cover more complex transactions between entities. Also, we do not limit the interactions to uni- or bi-action cases. In contrast, we consider an interaction as an "ensemble" of actions when it comprises of more than one action.

[11] In the form of a contract and its associated exchanges carried out between two or more parties in the form of buyers and sellers to trade an asset (or service) for payment (Yang et al., 2012).

Thanks to the recent progress in data analytics (Marx, 2013), it is possible to technically analyze and take into account all individual actions and interactions among actors in almost any human system. This motivated us to consider a 'discrete' approach to economy and economical systems in which individual actions, not the flows, define the models. Although, when smoothed along time, the discrete models would converge to their associated standard flow models, they could provide more insight in finer granularities and also could be used to analyze a system with high degree of 'asynchronicity' and inhomogeneity with respect to interactions along the spatial dimension. The core of the proposed modeling approach are the economic 'particles.'

A single action or a set of (flexible) collocated actions between entities are modeled by one or more 'action objects' imagined as particles. We call these particles "gauge econons." This notation makes a metaphoric association between these economic particles and *Gauge bosons* in physics, which are assumed to carry the interactions among other physical particles (Langacker, 2009). Every gauge econon has its own 'characteristics' expressed in terms of its time interval and also the assets being transferred. Assets are expressed in form of a 4-channel model as shown in Figure 2. This 4-channel approach allows us to consider all asset exchanges, not just money-based ones, in order to prevent any possibility of loopholes which could be misused to move footprint around and intentionally isolate it within a 'designated' entity with minimal money-related exchanges with the 'core' entities. To build the four channels, we start with the three major sectors in economy: i) Primary (raw material and food), ii) secondary (finished goods), and iii) tertiary (services) (Dietrich and Krüger, 2010; Fisher, 1939; Lavoie, 1971; Stijepic, 2011). It is worth noting that although 4-sector models also exist (Kenessey, 1987), they are different from our 4-channel model, and we will consider them in generalizations in the future. Also, there have been some discussions to consider information as another sector on top of service (Strømmen-Bakhtiar and Razavi, 2011; Yigitcanlar, 2011), but again with respect to our scope in this work that is to show the feasibility of a unified framework to estimate responsibilities of every actor, it would have minimal impact. However, we move one step further and separate the *banking* from the third major sector because money transfer is the main interaction type with the end users in many businesses, especially Telco systems. Therefore, we have 4 sector/exchange types to be considered: i) Raw material and food, ii) Finished goods, iii) Services, and iv) Money. Every exchange type is considered be self-normalized in order to prevent inter-influence among them. To do this, each asset type is considered with a weight 1 even if it has a *money* value of a fraction of *direct* money exchanges of an entity. To support this normalization, we argue that if a firm claims an exchange type is negligible to them and its normalized measure hurts them, they can simply eliminate it from their business; otherwise, if they cannot eliminate it, it means that it is not negligible. Another option would be to eliminate footprint via that exchange type, not the exchange itself, i.e., by sourcing it from an entity that has a lower footprint intensity. The ability to estimate the footprint and responsibility carried via the interactions of an entity with the other entities, directly or indirectly via other intermediating entities in between, is an important advantage of the MEIO models. Examples of the assets transferred along these four channels is provided in Table 1.

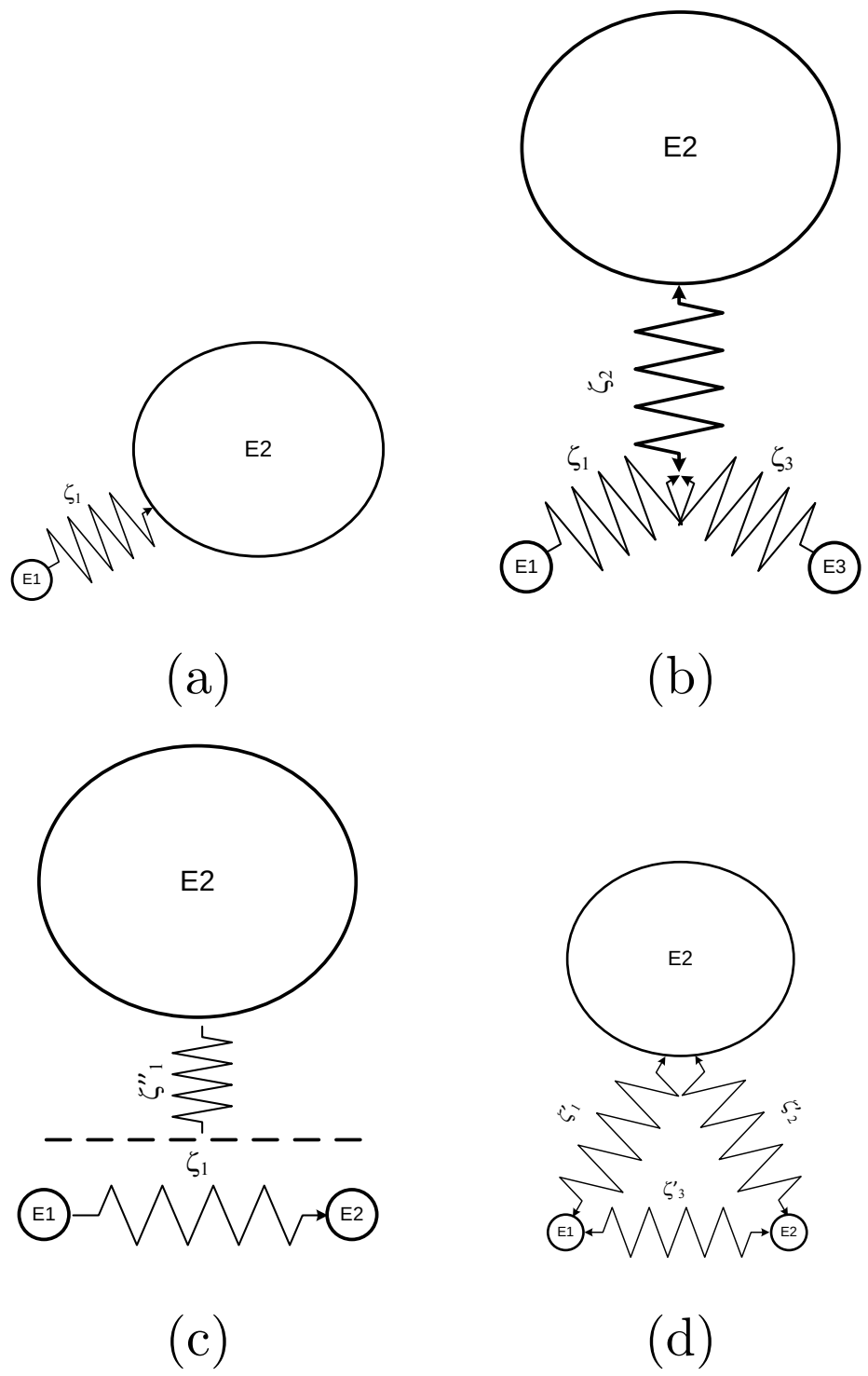

Figure B.1: Some basic gauge econons examples.

Some examples of this gauge econons are shown in Figure B.1. Figure B.1(a) shows a simple interaction of just one transfer or action. It can represent an action of giving a gift (bonus). Figure B.1(b) shows a more complicated interaction among three entities, such as an e-transfer of a check. Note that three gauge econons facilitate it in this figure. When the impact on the third entity, i.e., the bank, is null at the end, for example if the money stays in the same bank, the interaction can be equivalently remodeled as a single action, as shown in Figure B.1(c). However, caution should be exercised because many transfers and interactions are not free of cost. Finally, Figure B.1(d) shows another way to model the interaction of Figure B.1(b) using three gauge binary econons.

It is worth noting that the proposed formalism in its current form presented in this work is 'static', and it would require to be enhanced with dynamic modeling in order to enable it with predictive and forecasting capabilities (Lawson, 2007; Nguyen Huu and Costa-Lima, 2014). Paradigms adapted to individuals behavior, such as Fuzzy Cognitive Maps (FCMs) (Giabbanelli et al., 2014; Mago et al., 2012; Tsadiras and Bassiliades, 2014), will be considered in the future work in order to develop a framework for complex interaction systems with added behavior prediction capabilities.

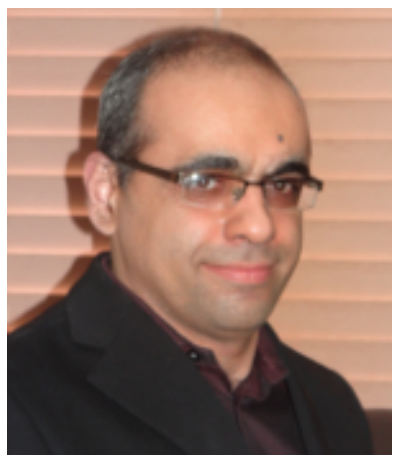

**Reza FARRAHI MOGHADDAM** received his B.Sc. degree in Electrical Engineering and his Ph.D. degree in Physics from the Shahid Bahonar University of Kerman, Iran, in 1995 and 2003, respectively. He has been a Postdoctoral Research Fellow and a Research Associate with the Synchromedia Laboratory for Multimedia Communication in Telepresence, École de technologie supérieure (ETS), University of Quebec, Montreal (QC), Canada since 2007 and 2012, respectively. Dr. Farrahi has published more than 50 technical papers, and his research interests include green computing, sustainable ICT, behavior analysis, life cycle analysis, green economy, image processing, visual perception, and optimization.

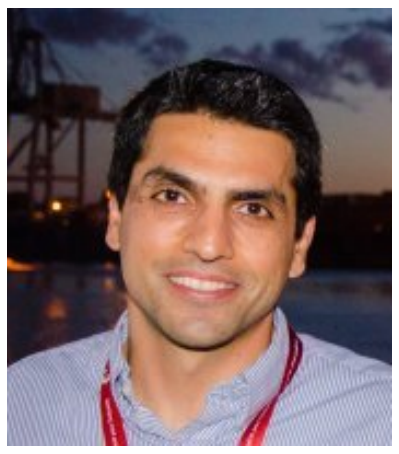

**Fereydoun FARRAHI MOGHADDAM** received his B.Sc. degree in Electronics Engineering from Shahid Bahonar University of Kerman, Iran, in 1999. He obtained his M.Sc. degree in Electronics Engineering from K.N. Toosi University of Technology, Tehran, Iran, in 2001. He joined the Synchromedia Laboratory for Multimedia Communication in Telepresence, École de technologie supérieure (ETS), University of Quebec, Montreal (QC), Canada as a Ph.D. student in September 2008, and received his Ph.D. degree on green distributed networks of data centers in 2014. He is currently working as a Postdoctoral Research Fellow with Synchromedia Lab since 2014. His research interests include green cloud computing, smart scheduling, and optimization.

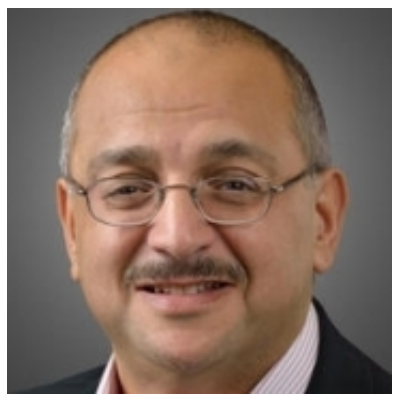

**Mohamed CHERIET** received his B.Eng. from USTHB University (Algiers) in 1984 and his M.Sc. and Ph.D. degrees in Computer Science from the University of Pierre et Marie Curie (Paris VI) in 1985 and 1988 respectively. Since 1992, he has been a professor in the Automation Engineering department at the École de technologie supérieure (University of Quebec), Montreal, and was appointed full professor there in 1998. His interests include document image analysis, OCR, mathematical models for image processing, pattern classification models and learning algorithms, as well as perception in computer vision. Dr. Cheriet has published more than 250 technical papers in the field, and has served as chair or co-chair of the following international conferences: VI'1998, VI'2000, IWFHR'2002, ICFHR'2008, and ISSPA'2012.

# A Multi-Entity Input Output (MEIO) Approach to Sustainability — Water-Energy-GHG (WEG) Footprint Statements in Use Cases from Auto and Telco Industries

Reza Farrahi Moghaddam[a,b,*], Fereydoun Farrahi Moghaddam[a,b], Mohamed Cheriet[a,b]

[a] *Synchromedia Lab, École de technologie supérieure (ETS), University of Quebec (UduQ), Montreal, QC, Canada*
[b] *CIRODD: P.O.Box 6079 Downtown, Montreal, QC, Canada, H3C 3A7*

---

**Abstract**

A new Input-Output model, called the Multi-Entity Input-Output (MEIO) model, is introduced to estimate the *responsibility* of entities and actors of an ecosystem on the footprint and actions of each other. It assumed that the ecosystem is comprised of end users, service providers, and utilities. The proposed MEIO modeling approach can be seen as a realization of the Everybody-in-the-Loop (EitL) framework, which promotes a sustainable future using behaviors and actions that are aware of their ubiquitous eco-socio-environment impacts. In this vision, the behavioral changes could be initiated by providing all actors with their footprint statement, which would be estimated using the MEIO models. First, a naive MEIO model is proposed in the form of a graph of actions and responsibility by considering interactions and goods transfers among the entities and actors along four channels. In this model, the unnormalized responsibility and also the final responsibility among the actors are introduced, and then are used to re-allocate immediate footprint of actors to those who are implicitly responsible. The footprint in the current model is limited to three major impacts: Water usage, Energy consumption, and GHG emissions. The naive model is then generalized to Provider-perspective (P-perspective) and End User-perspective (E-perspective) MEIO models in order to make it more suitable to cases where a large number of end users are served by a provider. The E-perspective modeling approach particularly allows estimating the footprint associated to a specific end user. In two use cases from the auto and Telco industries, it has been observed that the proposed MEIO models are practical and dependable in allocating footprint to the provider and also to the end user, while i) avoiding footprint leakage to the end users and ii) handling the large numbers end users. In addition, it will be shown that the MEIO models could implicitly provide some features of the Scope-3 and LCA approaches. This would make the MEIO models an interesting candidate for integrating and merging various concepts that are otherwise incompatible.

---

*Corresponding author: Reza Farrahi Moghaddam

*Email addresses:* `imriss@ieee.org` (Reza Farrahi Moghaddam), `farrahi@ieee.org` (Fereydoun Farrahi Moghaddam), `mohamed.cheriet@etsmtl.ca` (Mohamed Cheriet)

*URL:* `http://ca.linkedin.com/in/rezafm` (Reza Farrahi Moghaddam)

| | Steel | Al |
|---|---|---|
| Price ($/tonne) | 900.00 | 2,667.42 |
| Production (Mt) | 1,500.00 | 44.10 |
| Footprint Factor ($MtCO_2e$/Mt) | 2.82 | 12.00 |
| Footprint Factor ($tCO_2e$/MWh) | 0.76 | 0.76 |
| Footprint Factor (GWh/Mt) | 5,527.00 | 43,778.00 |
| Footprint Factor (GGal/Mt) | 18.00 | 8.31 |
| Total Footprint ($MtCO_2e$) | 10,530.78 | 1,996.46 |
| Total Footprint (GWh) | 8,290,500.00 | 1,930,609.80 |
| Total Footprint (GGal) | 27,000.00 | 366.47 |
| Honda Consumption (t) | 241,459.50 | 28,540.50 |
| Honda Payed ($M) | 217.31 | 76.13 |
| ROW Consumption (Mt) | 1,499.76 | 44.07 |
| ROW Payed ($M) | 1,349,782.69 | 117,557.09 |

Table S.1

## Supplementary Material S. Honda and BT Use cases

In this section, the details of the data used in sections 4.1 and 4.2 is provided.

### *Supplementary Material S.1. Data of the Honda Use case*

The set of EoI of this use case are: **Honda** The Honda Motor Co. Ltd, **Investors** The Honda's investors, **MatSteel** The Steel industry across the globe, **MatAl** The Aluminum industry across the globe, **UtlJP** The electricity utilities in Japan, **UtlNA** The electricity utilities in North America, **UtlSA** The electricity utilities in South America, **UtlEU** The electricity utilities in European Union, **UtlAO** The electricity utilities in Asia-Oceanic, and **UtlCH** The electricity utilities in China. Also, In terms of the reference market and prices, we use the NYSE market and the period of FY2011.

The two commodity used in this use case, i.e., Al and Steel, coincidentally have a peak price of $2,667.42 per metric ton and $900.00 per metric ton at the same time in April 2011.[1] The total worldwide Al production in 2011 was estimated 44,100 thousand metric tons, while the total raw steel production was estimated 1,500 million metric tons.[2] The average immediate footprint of Al production is estimated to 10.84 $KgCO_2$ per Kg of metal, and in the case of steel it is estimated to be 2.53 $KgCO_2$ per Kg of metal (Kissinger et al., 2013). In terms of energy consumption for steel production, it is estimated at 5,527 kWh/ton in Germany and 23,750 kWh in China.[3] In the case of Al, it is estimated at 43,778 kWh/ton in Germany.[4] The water consumption in Steel industry is estimated at 13,000-23,000 Gal/ton.[5], while it is estimated at 8,308.7 gal/ton for Al industry.[6] Also, an average of 0.76 $MtCO_2e$/MWh for indirect GHG emissions associated with the electricity consumption was assumed for Asia/Oceania region. Table S.1 summarizes the production, market, also footprint of the Al and Steel industries along with the share of Honda. The data of Honda was collected from Honda Environmental Annual Report 2011.[7]

[1] http://www.indexmundi.com/commodities/?commodity=aluminum&months=60&commodity=cold-rolled-steel
[2] http://minerals.usgs.gov/minerals/pubs/commodity/aluminum/mcs-2012-alumi.pdf
http://minerals.usgs.gov/minerals/pubs/commodity/iron_&_steel/mcs-2012-feste.pdf
[3] http://vfp1.vestforsk.no/sip/pdf/Felles/MetalProduction.pdf
[4] http://vfp1.vestforsk.no/sip/pdf/Felles/MetalProduction.pdf
[5] http://www.epa.gov/sectors/pdf/2008/iron_steel.pdf
[6] http://www.albuw.ait.ac.th/Groups/Assignment/II/Group-05pp.pdf
[7] http://world.honda.com/environment/report/download/pdf/2011_report_E_full.pdf

| | Japan | North America | South America[a] | Europe | Asia/Oceania[b] | China |
|---|---|---|---|---|---|---|
| Total Utility Footprint (TWh) | 997.40 | 4,701.07 | 584.66 | 3,635.60 | 1,346.42 | 3,503.40 |
| Total Utility Footprint ($MtCO_2e$) | 417.51 | 2,983.31 | 82.43 | 1,413.08 | 1,030.95 | 2,959.85 |
| Total Utility Footprint (Ggal water) | 508.67 | 2,209.52 | 280.64 | 1,490.60 | 511.63 | 1,961.90 |
| Utility Sell ($M) | 153,998.56 | 328,134.69 | 57,103.74 | 383,919.36 | 120,450.73 | 262,755.00 |
| Honda Electricity (TWh) | 1.60 | 1.54 | 0.33 | 0.15 | 0.95 | 0.58 |
| Honda Payed ($M) | 247.04 | 107.49 | 32.23 | 15.84 | 84.99 | 43.50 |
| ROW Electricity (TWh) | 995.80 | 4,699.53 | 584.33 | 3,635.45 | 1,345.47 | 3,502.82 |
| ROW Payed ($B) | 153.75 | 328.03 | 57.07 | 383.90 | 120.37 | 262.71 |

Table S.2: The electricity production and footprint of utilities and their relations with Honda in 2011. [a] Only Brazil, Argentina, and Colombia are considered in the calculations. [b] Only India, New Zealand, Vietnam, Pakistan, Philippines, Malaysia, Turkey, and Thailand are considered in the calculations.

Honda electricity consumption and its interactions with various utilities across the world are provided in Table S.2.[8] The details of calculating the footprint factors of each utility region is presented in Table S.3. For each region, as provided in Table S.4, the average electricity mix is used to calculated the associated water footprint factors.[9] It is worth mentioning the associated Global Warming Potentials (GWPs) on a 100-year time horizon: $\mathrm{GWP}_{\mathrm{CO_2}} = 1$, $\mathrm{GWP}_{\mathrm{CH_4}} = 25$, $\mathrm{GWP}_{\mathrm{N_2O}} = 298$.[10] Also, the associated market volume and pricing of the utilities are provided in Table S.5.[11]

The Honda's footprint summary is presented in Table S.6.[12] The data is from Honda environmental reports and also the calculations and results of aforepresented tables. As can be seen from the table, the water footprint associated to electricity production was not included in the final sum because of having doubt if it has been already included in the reported water consumption. Therefore, we crossed this figure out in order to avoid any possible double counting. The same policy was used for the GHG emissions footprint, and only the reported figure of 1,240.00 $KtCO_2e$ was used.

The Honda's sell in FY2011 and FY2012 is presented in Table S.7. As mentioned before, we intentionally use the Scope-3 reported footprint of the buyers in FY2011 for FY2012 in order to make our results

[8] http://world.honda.com/environment/report/download/pdf/2011_report_E_full.pdf
http://www.iea.org/publications/freepublications/publication/key_world_energy_stats.pdf
http://www.world-nuclear.org/why/greenhouse_gas_from_generation.html
http://en.wikipedia.org/wiki/Electric_energy_consumption
http://www.eia.gov/oiaf/1605/pdf/Appendix\%20F_r071023.pdf

[9] http://www.nrel.gov/docs/fy11osti/50900.pdf
http://www.eia.gov/countries/analysisbriefs/Japan/japan.pdf
http://www.energy.eu/publications/KOAE09001_002.pdf
http://www-pub.iaea.org/mtcd/publications/pdf/pub1304_web.pdf

[10] IPCC's Fourth Assessment Report (2007) and Fifth Assessment Report (2013):
http://www.ipcc.ch/pdf/assessment-report/ar4/wg1/ar4-wg1-chapter2.pdf,
http://www.climatechange2013.org/images/report/WG1AR5_Chapter02_FINAL.pdf.

[11] http://web.ing.puc.cl/~power/paperspdf/RudnickIEE.pdf
http://www.powerexindia.com/PXIL/
http://userpage.fu-berlin.de/~ballou/fama/vietnam/vnelectricity.pdf
http://www.lesco.gov.pk/CustomerServices/3000063.asp
http://www.demotix.com/news/1153471/philippines-has-highest-electricity-costs-asia#media-1151730
http://en.wikipedia.org/wiki/Electricity_sector_in_Colombia
http://en.wikipedia.org/wiki/Electricity_pricing

[12] http://www.ec.gc.ca/ges-ghg/default.asp?lang=En&n=AC2B7641-1
http://astro.berkeley.edu/~wright/fuel_energy.html

| | $CO_2$ Factor (Kg/MWh) | $CH_4$ Factor (g/MWh) | $N_2O$ Factor (g/MWh) | GHG Factor (Kge/MWh) | Total Electricity (TWh) | Total GHG Footprint ($MtCO_2e$) |
|---|---|---|---|---|---|---|
| **Japan** | 417.00 | 8.39 | 4.65 | 418.60 | 997.40 | 417.51 |
| USA | 687.00 | **20.51** | **11.36** | | 3,961.56 | |
| Canada | 223.00 | 3.90 | 3.51 | | 521.85 | |
| Mexico | 593.00 | 16.76 | 2.30 | | 217.66 | |
| **NA** | 631.14 | 18.49 | 10.07 | 634.60 | 4,701.07 | 2,983.31 |
| Brazil | 93.00 | 2.51 | 1.06 | | 426.34 | |
| Argentina | 317.00 | 5.70 | 1.01 | | 110.52 | |
| Colombia | 157.00 | 2.80 | 1.85 | | 47.80 | |
| **SA** | 140.58 | 3.14 | 1.12 | 140.99 | 584.66 | 82.43 |
| **EU** | 387.00 | 6.94 | 5.05 | 388.68 | 3,635.60 | 1,413.08 |
| India | **999.00** | 16.64 | **19.59** | | 689.54 | |
| New Zealand | 159.00 | 3.07 | 0.84 | | 40.34 | |
| Vietnam | 417.00 | 12.97 | 3.89 | | 78.93 | |
| Pakistan | 482.00 | **31.46** | 5.49 | | 76.61 | |
| Philippines | 526.00 | 15.54 | 7.77 | | 54.42 | |
| Malaysia | 528.00 | 19.84 | 3.65 | | 101.00 | |
| Turkey | 584.00 | 11.35 | 6.28 | | 165.09 | |
| Thailand | 583.00 | 19.67 | 4.89 | | 140.49 | |
| **Asia/Oceania** | 761.56 | 16.72 | 12.47 | 765.69 | 1,346.42 | 1,030.95 |
| **China** | 839.00 | 14.58 | **18.41** | 844.85 | 3,503.40 | 2,959.85 |

Table S.3: The breakdown of utilities at the nation level across six regions considered in this work.

| | Gas | Oil | Coal | Nuclear | Hydro | Renewables | Overall Water Factor (Gal/kWh) | Total Electricity (TWh) | Total Water Footprint (Ggal) |
|---|---|---|---|---|---|---|---|---|---|
| Water Factors (Gal/kWh) | 0.43 | 0.20 | 0.72 | 0.72 | 3.25 | 0.03 | | | |
| Japan (%) | 18.00 | 42.00 | 22.00 | 13.00 | 3.00 | 0.00 | | | |
| Japan | | | | | | | 0.51 | 997.40 | 508.67 |
| NA (%) | 24.30 | 39.80 | 22.00 | 8.40 | 2.10 | 0.50 | | | |
| NA | | | | | | | 0.47 | 4,701.07 | 2,209.50 |
| SA (%) | 24.60 | 45.70 | **3.70** | 0.90 | **7.60** | 1.00 | | | |
| SA | | | | | | | 0.48 | 584.66 | 280.64 |
| EU (%) | 23.90 | 36.40 | 18.30 | 13.40 | 0.20 | **7.80** | | | |
| EU | | | | | | | 0.41 | 3,635.60 | 1,490.60 |
| Asia/Oceania (%) | 24.05 | 40.45 | 21.40 | 0.20 | 1.15 | 0.50 | | | |
| Asia/Oceania | | | | | | | 0.38 | 1,346.42 | 511.64 |
| China (%) | 8.70 | 28.80 | **51.20** | 5.40 | 1.90 | 0.50 | | | |
| China | | | | | | | 0.56 | 3,503.40 | 1,961.90 |

Table S.4: The water footprint factors of the utilities. The fuel mix corresponds to 2005. Note: Biomass fuel type is ignored in this table.

| | Electricity Price ($/MWh) | Total Electricity (TWh) | Total Electricity Price ($M) |
|---|---|---|---|
| **Japan** | 154.40 | 997.40 | 153,998.56 |
| USA | 67.90 | 3,961.56 | |
| Canada | 69.90 | 521.85 | |
| Mexico | 104.20 | 217.66 | |
| **NA** | 69.80 | 4,701.07 | 328,134.69 |
| Brazil | 113.00 | 426.34 | |
| Argentina | 38.60 | 110.52 | |
| Colombia | 97.50 | 47.80 | |
| **SA** | 97.67 | 584.66 | 57,103.74 |
| **EU** | 105.60 | 3,635.60 | 383,919.36 |
| India | 73.20 | 689.54 | |
| New Zealand | 181.50 | 40.34 | |
| Vietnam | 58.82 | 78.93 | |
| Pakistan | 78.94 | 76.61 | |
| Philippines | 304.60 | 54.42 | |
| Malaysia | 74.20 | 101.00 | |
| Turkey | 131.00 | 165.09 | |
| Thailand | 44.60 | 140.49 | |
| **Asia/Oceania** | 89.46 | 1,346.42 | 120,450.73 |
| **China** | 75.00 | 3,503.40 | 262,755.00 |

Table S.5: The electricity production and price across the regions considered in this use case. Please note that the electricity price in France is used for the EU. Also, the household price is used for New Zealand.

| | Natural Gas | Liquefied Petroleum Gas (LPG) | Diesel | Total Footprint |
|---|---|---|---|---|
| GHG Factor ($KtCO_2e$/GLit) | 1.89 | 2,734.40 | 2,785.50 | |
| Water Factor (Gal/kWh) | 0.20 | 0.72 | 0.72 | |
| Conversion Factor (GJ/MLit) | 39.00 | 25,700.00 | 39,600.00 | |
| Honda Consumption (GJ) | 8,112,726.00 | 3,426,524.00 | 1,334,611.00 | |
| Honda Footprint Total (GWh) | 2,253.72 | 951.89 | 370.75 | 3,576.36 |
| Honda Footprint Total ($KtCO_2e$) | 393.16 | 364.57 | 93.88 | ~~851.61~~ |
| Honda Footprint Scopes 1 ($KtCO_2e$) | | | | 1,240.00 |
| Honda Footprint Elect (KGal) | 450.74 | 685.36 | 266.94 | ~~1,403.04~~ |
| Honda Footprint Tap (KGal) | | | | 4,165.20 |
| Honda Footprint Ground (KGal) | | | | 3,466.20 |
| Honda Footprint Total (KGal) | | | | 7,631.40 |

Table S.6: The summary of Honda's footprint in this use case. The water consumption for generating electricity is not summed up with other because we were uncertain about double counting.

| | Vehicles | Motorcycles | GHG per Vehicle ($tCO_2e$) | Total GHG ($MtCO_2e$) | Gasoline Factor ($kWh/tCO_2e$) | Energy per Vehicle (MWh) | Total Energy (TWh) |
|---|---|---|---|---|---|---|---|
| World (FY2012) | 3,117,000.00 | 15,284,000.00 | 18.21 | ←195.88 | 4,150.80 | 75.57 | 813.06 |
| World (FY2011) | 3,559,000.00 | 18,626,000.00 | 18.21 | 234.40 | 4,150.80 | 75.57 | 972.95 |

Table S.7: For 15 years starting from April 2010, (FY 2011). For the first line, we use the FY2012 as provided by Honda Scope-3 report. For both cars and motorcycles.

| Actor | Water Footprint (Mgal) | Energy Footprint (GWh) | GHG Footprint ($MtCO_2e$) |
|---|---|---|---|
| Honda | 7.63 | 3,576.36 | 1.24 |
| Buyers | 0.00 | 972,950.00 | 234.40 |
| Investors | 0.00 | 0.00 | 0.00 |
| MatSteel | 27,000,000.00 | 8,290,500.00 | 10,530.78 |
| MatAl | 366,470.00 | 1,930,610.00 | 1,996.46 |
| UtlJP | 508,670.00 | 997,400.00 | 417.51 |
| UtlNA | 2,209,520.00 | 4,701,070.00 | 2,983.31 |
| UtlSA | 280,640.00 | 584,660.00 | 82.43 |
| UtlEU | 1,490,600.00 | 3,635,600.00 | 1,413.08 |
| UtlAO | 511,630.00 | 1,346,420.00 | 1,030.95 |
| UtlCH | 1,961,900.00 | 3,503,400.00 | 2,959.85 |

Table S.8: The footprint of various entities involved in the Honda use case.

incomparable with actual numbers.[13] This has resulted in a footprint of 234.40 $MtCO_2e$ associated to the FY2011 buyers in our calculations. However, still our GHG emissions associated to a car, i.e., 18.21 $tCO_2e$, is very close to that reported by Honda (16.5 $tCO_2e$).[14] Also, the energy consumption associated to buyers' use of vehicles is calculated in the same table. It is worth mentioning that we ignored the black carbon footprint in this use case, which could be important especially from the air pollution point of view.

Table S.8 summarizes the footprint of all entities of this use case. Also, the monetary transfers among the entities, especially Honda, were retrieved from the Honda Consolidated Financial Results[15] Also, we assumed that the buyers pay for the vehicle's price in a period of 3 years, and also an average income of $50K was considered. The details of all gauge econons of this use case is provided in Table 6.

## *Supplementary Material S.2. Data of the BT Use case*

As mentioned in the text, in this use case the focus is to determine how the footprint of the provider is distributed among the end users and other entities. Therefore, the footprint of the end users is not considered, and it will be addressed in another study. In addition to the end users, investors and advertisement agencies are other important mobilizer entities, and therefore they should receive a considerable share of responsibility and footprint. This use case provides a way to achieve such a share among these high different entities of various sizes while being fair. Another interesting aspect of this use case is presence of several end user entities in the picture that shows the capability of the MEIO

[13]The footprint of buyers is calculated for a period of 15 years after purchase of the vehicle. Also, it is assumed that a motorcycle has as half as a car's footprint.

[14]`http://world.honda.com/news/2012/c120825Greenhouse-Gas-Emissions/index.html`
`http://www.honda.co.jp/environment/face/2012/case19/episode/episode06.html`
`http://astro.berkeley.edu/~wright/fuel_energy.html`

[15]`http://world.honda.com/investors/library/financialresult/2010/Financial_Result_2010_4q_E.pdf`

| | UK |
|---|---|
| Total Utility Footprint (TWh) | 351.80 |
| Total Utility Footprint ($MtCO_2e$) | 417.51 |
| Total Utility Footprint (Ggal water) | 186.45 |
| Utility Sell (M£) | 26,596.08 |
| BTUK Electricity (TWh) | 2.34 |
| BTUK Payed (M£) | 250.00 |
| ROW Electricity (TWh) | 349.46 |
| ROW Payed (B£) | 26.35 |

Table S.9: The UtlUK production and footprint in FY2011.

modeling in handling such cases.

We have chosen the case of BT as our use case here. However, because of limited available data, we down scaled the study to BT UK. The details of the data used in provided in the following section.

In addition, we acknowledge that this provider has contributed to sustainability in many other ways that are not considered in this use case, such as **1.** Charity and **2.** Contribution to Research. The authors plan to perform comprehensive analyses with the help of the providers to assess the significance of such actions in the future.

*Supplementary Material S.2.1. BT UK Use case - Data*

The union set of the EoI also the three end user entities of this use case is: **BTUK** The BT UK, **ClientsBB** The broadband-service clients, **ClientsPh** The phone-service clients, **ClientsTV** The TV-service clients, **Investors** The BT's investors, **Community** The community received contribution from BT, **UtlUK** The electricity utilities in UK, **Ads** The advertisement agencies. As can be seen, there are three entities that represent the end users.[16] Also, a period of one month is considered as $\Delta t$.

The summary of the UtlUK production and footprint in addition to BTUK share of electricity consumption is provided in Table S.9.[17] Also, the footprint factors of the UtlUK are retrieved from Energy Information Administration (EIA)'s Electricity Emission Factors,[18] and the GHG footprint is recalculated in Table S.10 which has a good agreement with the figure reported in Table S.9.

The performance of the UtlUK in terms of revenue and also water consumption is provided in Tables S.11 and S.12.[19] Finally, the summary of the entities' footprint is provided in Table S.13.

In terms of monetary transfers, total BTUK revenue was &M20,076.00 (with a profit before tax of &M1,717), from which &M543 was transferred to dividend investors. Also, BT UK had a pool of approximately 5, 394, 000 broadband clients[20] with an estimated plan of &52.36 per month.[21] Therefore, the total volume of channel four transfer between BTUK and ClientsBB would be &M338.96 per month. Also, the broadband clients had an average of 17GB per month per client data usage, which translates into a 9,174.22TB per month data usage for this entity on the channel three.

[16]For the purpose of simplicity, the overlapping cases where an end user is subscribed to two or more services are ignored. Also, entities such as Above.net, which perform peering, i.e., Settlement-free interconnection (Dhamdhere and Dovrolis, 2011), with BT, are ignored.

[17]`http://www.globalservices.bt.com/static/assets/pdf/case_studies/EN_NEW/bt_energy_efficiency_case_study.pdf`
`https://www.btplc.com/Sharesandperformance/Annualreportandreview/pdf/BTAnnualReport2012.pdf`
`http://www.iea.org/publications/freepublications/publication/key_world_energy_stats.pdf`
`http://www.eia.gov/countries/analysisbriefs/United_Kingdom/uk.pdf`

[18]`http://www.eia.gov/oiaf/1605/pdf/Appendix\%20F_r071023.pdf`

[19]The same sources as those of Table S.4 are used here.

| | $CO_2$ Factor[(a)] (Kg/MWh) | $CH_4$ Factor (g/MWh) | $N_2O$ Factor (g/MWh) | Ind. GHG Factor[(b)] (Kge/MWh) | GHG Factor (Kge/MWh) | Total Electricity (TWh) | Total GHG Footprint ($MtCO_2e$) |
|---|---|---|---|---|---|---|---|
| **UtlUK** | 475.00 | 7.93 | 5.49 | 701.00 | 1,177.83 | 351.80 | 414.36 |

Table S.10: The GHG footprint of UtlUK calculated using its footprint factors. It is in close agreement with the reported GHG footprint of the UK utility, i.e., 417.51 $MtCO_2e$. a) Emission inventory electricity emission factors are based on average emissions intensity of total electric sector generation for UK and include transmission and distribution (T&D) losses incurred in delivering electricity to the point of use. b) Indirect emission reductions emission factors for reduced purchases of electricity are based on average emissions intensity of fossil-fired generation for UK and include transmission and distribution (T&D) losses incurred in delivering electricity to point of use.

| | Electricity Price (£/MWh) | Total Electricity (TWh) | Total Electricity Price (M£) |
|---|---|---|---|
| **UtlUK** | 75.60 | 351.80 | 26,596.08 |

Table S.11: The total production and revenue of UtlUK.

| | Gas | Oil | Coal | Nuclear | Hydro | Renewables | Overall Water Factor (Gal/kWh) | Total Electricity (TWh) | Total Water Footprint (Ggal) |
|---|---|---|---|---|---|---|---|---|---|
| Water Factors (Gal/kWh) | 0.43 | 0.20 | 0.72 | 0.72 | 3.25 | 0.03 | | | |
| UtlUK (%) | 46.00 | 2.00 | 28.00 | 13.00 | 1.00 | 6.00 | | | |
| UtlUK | | | | | | | 0.53 | 351.80 | 186.45 |

Table S.12: The performance of the UtlUK in terms of water consumption.

| Actor | Water Footprint (Mgal) | Energy Footprint (GWh) | GHG Footprint ($ktCO_2e$) |
|---|---|---|---|
| BTUK | 30.33 | 195.03 | 52.83 |
| ClientsBB | 61.39 | 115.83 | 137.51 |
| ClientsPh | 2.20 | 4.16 | 4.94 |
| ClientsTV | 3.36 | 6.34 | 7.85 |
| Investors | 0.00 | 0.00 | 0.00 |
| Community | 0.00 | 0.00 | 0.00 |
| UtlUK | 15,537.50 | 29,316.67 | 34,790.00 |
| Ads | 0.00 | 0.00 | 0.00 |

Table S.13: The footprint of various entities involved in the BT use case for FY2011.

For the case of phone clients, the total revenue was &M4,714 from 17.02 Million subscribers. The ClientsPh used 4,152.83M minutes of service per month with an estimated plan rate of &23.08.[22]

Finally, the TV users, the ClientsTV, were 4% of the UK's TV subscribers (572K clients). The total BTUK revenue was &M144.8 from advertisement, and &M347.2 from subscribers (with an estimated plan of &50.58 per month). We can estimate a monetary transfer of &B3.62 from advertisement agencies to the UK's TV providers.[23]

For all clients, we assume an average yearly income of &26,500. Therefore, the monetary transfer from ClientsBB, ClientsPh, and ClientsTV to BTUK is 2.36%, 1.05%, and 2.29% share of their channel four transfers, respectively.

In terms of clients footprint, we use very simple approximations. For the ClientsBB, we consider they use BT Home Hub (3.0) for 5 years. It has been reported by the provider that this equipment has a LCA footprint of 164 Kg$CO_2$e (92% in-use and 8% embodied) (BT, 2012). This is equal to 2.52 Kg$CO_2$e per month footprint for the hub. Also, we assume a client uses a laptop for access with electricity consumption of $2.34 per month, which is equal to 19.5 kWh electricity consumption per month. For the UK's grid mix, this means a client also has a footprint of 9.30 Kg$CO_2$e because of laptop energy consumption. Adding these two figures and considering the total number of the ClientsBB members, we obtain a footprint of 63.75 Kt$CO_2$e per month for this entity. The associated energy and water consumption footprint would be 131,445.8 MWh and 69.667 kgal, respectively.

For the phone clients, we assume they use BT Graphite 2500 DECT phone for the same period of 5 years. This phone has 23 Kg$CO_2$e (75% in-use and 25% embodied) (BT, 2012). For the 17.02 million of ClientsPh, this is equal to 4.94 Kt$CO_2$e per month for this entity. The energy and water footprint would be 10,263.85 MWh and 5.44 kgal, respectively.

Finally, for the ClientsTV entity, we assume a client uses BT Vision+ digital settop box for 4 years with a footprint of 354 Kg$CO_2$e (87% in-use and 13% embodied) (BT, 2012). This is equal to a footprint of 5.13 Kg$CO_2$e per month. Also, we assume they use a 60W TV set for 4 hours a day. This is equal to 7.3 kWh electricity consumption (i.e., 3.48 Kg$CO_2$e because of TV set). The total footprint of a client would be 8.61 Kg$CO_2$e, 18.06 kWh, and 9.57 gal per month respectively. For the total size of 527 thousands clients on the ClientsTV, we obtain a footprint of 4.54 kt$CO_2$e, 9,519.22 MWh, and 5.05 kgal per month, respectively. We also assume no immediate footprint for investors, community, and advertisement agencies.

---

[20] It accounts for 28.2% of all its clients: `http://www.ispreview.co.uk/story/2011/02/03/uk-isp-bt-retail-tops-5-5-million-broadband-customers.html`
`http://stakeholders.ofcom.org.uk/binaries/research/cmr/telecoms/q4-2011.pdf`
`http://www.guardian.co.uk/technology/2011/nov/01/home-broadband-download-17-gigabytes`

[21]

[22] `http://stakeholders.ofcom.org.uk/binaries/research/cmr/telecoms/q4-2011.pdf`

[23] The total revenue of all UK TV providers was &B12.3.